\pdfoutput=1
\documentclass[fleqn,usenatbib]{mnras}
\usepackage[T1]{fontenc}

\DeclareRobustCommand{\VAN}[3]{#2}
\let\VANthebibliography\thebibliography
\def\thebibliography{\DeclareRobustCommand{\VAN}[3]{##3}\VANthebibliography}

\usepackage{graphicx}
\usepackage{amsmath}
\usepackage[normalem]{ulem}

\usepackage{sidecap}
\sidecaptionvpos{figure}{t}

\usepackage{ae,aecompl}
\usepackage{url}
\usepackage{amssymb}

\usepackage{booktabs}
\usepackage{etoolbox}
\usepackage{relsize}
\usepackage{multirow}
\usepackage{array}
\usepackage{amsfonts}
\usepackage{gensymb}

\usepackage{color}
\usepackage{multirow}
\usepackage{multicol}
\usepackage{siunitx}
\usepackage{threeparttable}
\usepackage{verbatim}
\usepackage{float}
\usepackage{booktabs}
\usepackage{dcolumn}
\usepackage{algorithm,algpseudocode}
\hypersetup{colorlinks=true,linkcolor=red,citecolor=blue,breaklinks=true}
\usepackage[english]{babel}
\usepackage{array}
\usepackage[capitalise,noabbrev]{cleveref}
\usepackage{upgreek}
\usepackage{orcidlink}

\usepackage{adjustbox}
\usepackage{makecell}

\def\deg{\ifmmode^\circ\else$^\circ$\fi}

\newcommand{\dd}{\mathrm{d}} 
\newcommand{\e}{\mathrm{e}} 

\newcommand{\DP}{\text{DP}}     
\newcommand{\PI}{\text{PI}}     
\newcommand{\RM}{\text{RM}}     
\newcommand{\el}{_\text{e}}     
\renewcommand{\vec}[1]{\boldsymbol{#1}}
\newcommand{\cm}{\,\text{cm}}    
\newcommand{\m}{\,\text{m}}      
\newcommand{\p}{\,\text{pc}}     
\newcommand{\kpc}{\,\text{kpc}}  

\newcommand{\mkG}{\,\upmu\text{G}} 

\newcommand{\rad}{\,\text{rad}} 
\newcommand{\radm}{\rad\m^{-2}}

\RequirePackage{mymacro}
\graphicspath{{./Figures/}}

\title[Magnetic field in M31]{The fine structure of the mean magnetic field in M31}

\author[Paul et al.]{Indrajit Paul,$^{1}$\thanks{E-mail: indrajit.paul@niser.ac.in (IP), vasanth.kashyap@niser.ac.in (RVK), tghosh@niser.ac.in (TG), rbeck@mpifr-bonn.mpg.de (RB), sinha.srijita@niser.ac.in (SS), anvar.shukurov@ncl.ac.uk (AS)}\orcidlink{0009-0002-2521-0074}
R.~Vasanth Kashyap,$^{1}$\orcidlink{0009-0007-7236-1808}
Tuhin Ghosh,$^{1}$\orcidlink{0000-0001-6088-3034}
Rainer Beck,$^{2}$\orcidlink{0009-0001-8154-3562}
\newauthor
Luke Chamandy,$^{1}$\orcidlink{0000-0003-4935-5550}
Srijita Sinha, $^{1}$\orcidlink{0000-0002-4805-3654}
and Anvar Shukurov$^{3}$\orcidlink{0000-0001-6200-4304}
\\
$^{1}$National Institute of Science Education and Research, An OCC of Homi Bhabha National Institute, Bhubaneswar 752050, Odisha, India\\
$^{2}$Max-Planck-Institut f\"ur Radioastronomie, Auf dem H\"ugel 69, 53121 Bonn, Germany\\
$^{3}$School of Mathematics, Statistics and Physics, Newcastle University, Newcastle upon Tyne, NE1 7RU, UK
}

\date{Accepted XXX. Received YYY; in original form ZZZ}

\pubyear{2025}

\begin{document}

\label{firstpage}
\pagerange{\pageref{firstpage}--\pageref{lastpage}}
\maketitle

\begin{abstract}
To explore the spatial variations of the regular (mean) magnetic field of the Andromeda galaxy (M31), we use Fourier analysis in azimuthal angle along four rings in the galaxy's plane. The Fourier coefficients give a quantitative measure of strength of the modes, enabling us to compare  expectations from mean-field dynamo models of spiral galaxies. Earlier analyses indicated that the axisymmetric magnetic field  (azimuthal Fourier mode $m=0$) is sufficient to fit the observed polarization angles in a wide range of galactocentric distances. We apply a Bayesian inference approach to new, more sensitive radio continuum data at $\lambda \lambda3.59$, $6.18$, and $11.33\cm$ and the earlier  data at $\lambda 20.46\cm$ to reveal sub-dominant contributions from the modes $m=1$, 2, and 3 along with a dominant axisymmetric mode. Magnetic lines of the axisymmetric mode are close to trailing logarithmic spirals which are significantly more open than the spiral arms detectable in the interstellar dust and neutral hydrogen. The form of the $m=0$ mode is consistent with galactic dynamo theory. Both the amplitudes and the pitch angles of the higher azimuthal modes ($m>1$) vary irregularly with $r$ reflecting local variations in the magnetic field structure. The maximum strength of the mean magnetic field of $1.8\text{--}2.7\mkG$ (for the axisymmetric part of the field) occurs at $10\text{--}14\kpc$ but we find that its strength varies strongly along the azimuth; this variation gives rise to the $m=1$ mode. 
We suggest a procedure of Bayesian inference which is independent of the specific nature of the depolarization and applies when the magneto-ionic layer observable in polarized emission is not symmetric along the line of sight because emission from its far side is completely depolarized.
\end{abstract}

\begin{keywords}
galaxies: individual: M31 -- galaxies: spiral -- galaxies: magnetic field -- ISM: magnetic fields --  radio continuum: galaxies
\end{keywords}

\section{Introduction} \label{sec:intro}

Magnetic fields play a crucial and multifaceted role in astrophysics, impacting various processes and phenomena in galaxies. In particular, they affect the propagation of cosmic rays \citep{zweibel2013microphysics}, the formation and evolution of galaxies \citep{imagine18,tabatabaei2018role}, the star formation rate, and the initial mass function \citep{10.3389/fspas.000072019.}. A reliable, detailed knowledge of the structure of a large-scale galactic magnetic field is essential for the mean-field dynamo theory (for reviews, see \citealt{Beck:2019} and \citealt{,SS21}).

Magnetic fields can exist on small turbulent scales or as large-scale coherent structures, depending on the physical processes involved. Polarized synchrotron emission enables the study of the ordered component of the magnetic field projected onto the sky plane using radio observations. Galactic magnetic fields appear more ordered in radio observations than in the far-infrared  \citep{borlaff:2023}, suggesting that different parts of the interstellar medium and different scales are probed at different wavelength ranges.

Magnetic fields ordered on a kiloparsec scale in spiral galaxies have two contributions: regular (mean) fields, which vary little in either direction or strength, and anisotropic random fields, which exhibit frequent field reversals but a similar anisotropic orientation \citep{2015A&ARv..24....4B}. Both types of the magnetic field contribute to the polarized radio emission. The Faraday rotation of the polarization angle, combined with the density of thermal electrons, provides a method to determine both the strength and direction of the line-of-sight component of the regular magnetic field and distinguish between the regular and anisotropic random interstellar magnetic fields.

Although random fields do not contribute to the mean Faraday rotation, they contribute to its dispersion. Randomly oriented small-scale turbulent fields decrease the degree of polarization, especially at longer wavelengths, via a range of mechanisms collectively known as Faraday depolarization \citep{burn1966depolarization,sokoloff1998depolarization}.
Various configurations of synchrotron-emitting regions (e.g., electron density and magnetic field strength distributions) can also affect the depolarization. By observing polarized synchrotron emission over a sufficiently wide range of wavelengths, it is possible to investigate both the depolarization effects and the Faraday rotation, providing insights into the magnetic field structure and the intervening medium. If the anisotropic random field is negligible compared to the regular field,  it becomes possible to infer the three-dimensional (3D) global magnetic field structure of the galaxy by combining polarized synchrotron and Faraday rotation data using multi-frequency radio observations.

Depolarization mechanisms in the radio range can be categorized as either wavelength-independent or wavelength-dependent \citep{burn1966depolarization,sokoloff1998depolarization}. Wavelength-independent depolarization is typically caused by the tangling of the magnetic field within the telescope beam. In contrast, wavelength-dependent depolarization arises from factors such as rotation measure (RM) gradients, differential Faraday rotation, and both internal and external Faraday dispersion. The wavelength-dependent change in the polarization between the wavelengths $\lambda_1$ and $\lambda_2$ ($\DP_{{\lambda_1}/{\lambda_2}}$) is quantified in terms of the ratio of polarization intensities at those wavelengths. When magnetic field fluctuations occur within a region with both synchrotron emission and Faraday rotation, the dominant cause of depolarization is usually the internal Faraday dispersion \citep{williams2024disentangling}.

The large-scale magnetic field structures in nearby galaxies have been extensively studied using radio polarimetric observations in the wavelength range $3$ to $20\cm$ (see table~4 of \citealt{Beck:2019}). The polarization angle ($\psi$) is measured in the sky plane with respect to the local outward radial direction in the galactic mid-plane. A model of the regular magnetic field can capture the azimuthal variation of $\psi$ (corrected for the RM) with a set of Fourier modes ($m$) in local cylindrical coordinates \citep{Sokolov:1992, BHKNPSS97}.
For example, the regular field of M51 has been presented as a combination of axisymmetric ($m=0$) and quadrisymmetric ($m=2$) modes across four  rings at galactocentric distances 2--7\,kpc \citep{fletcher2011magnetic}. In the case of M33, \cite{tabatabaei2008high} modelled the regular field in two radial ranges (1--5\,kpc) and found a statistically good fit for multiple Fourier mode combinations mostly  dominated by the $m=0$ mode. Results of the mode analysis for further spiral galaxies are given in Table~5 of \citet{Beck:2019}.
These analyses are consistent with the dominance of the axisymmetric large-scale magnetic structures ($m=0$) affected by a spiral pattern (producing weaker modes with $m>0$), except for the barred galaxies where strong deviations from axial symmetry in the velocity field and the gas density are reflected in a more complicated magnetic structure.

The Andromeda nebula, M31 or NGC~224, the nearest spiral galaxy to the Milky Way, has been extensively studied, including its magnetic field structure. We use the distance of $780\kpc$ ($1\arcmin\approx230\p$) and $i=75^\circ$ in this paper \citep{BBGM20}. Mainly because of its close distance and large angular size, we can observe M31 with a spatial resolution of better than $170\p$, which is 45\arcsec\ in angular scale. 
The continuum synchrotron radiation is mostly concentrated in a ring at a radius of about $10\kpc$ in the galaxy plane \citep{1977A&A....57....9B}. \cite{Fletcher:2004} applied the azimuthal Fourier mode analysis to M31 using the radio polarization observations available at that time at the wavelengths $6.18\cm$ \citep{1982A&A...106..121B},  $11.33\cm$ \citep{berkhuijsen2003polarized}, and $20.46\cm$ \citep{1982A&A...106..121B}, assuming a distance of $690\kpc$  \citep{Baade:1963}  and inclination angle $i=78\deg$ \citep{Braun91} ($i=0\deg$ corresponds to the face-on view). Ring-wise analysis of the observed polarization angle variation with the azimuthal angle revealed the presence of a dominant $m=0$ mode in the regular field for the outer three rings (8--14\,kpc). The innermost ring (6--8\,kpc) required an additional contribution of the $m=2$ mode along with the dominant $m=0$ mode. \cite{BBGM20} analysed the latest radio polarization data from the Effelsberg $100$-m telescope at $\lambda = 3.59, \ 6.18$, and $11.33\cm$ using an updated distance measurement of  $780\kpc$ \citep{stanek1998distance} and $i=75\deg$ \citep{chemin2009h}. The azimuthal variation of Faraday rotation measure indicates the presence of the Fourier modes $m=0$ and $m=1$ in the regular magnetic field. The presence of $m=1$ mode in the analysis of \cite{BBGM20} is attributed to the use of the updated radio observations of M31 and the Fourier mode analysis of Faraday rotation measure data alone, rather than the total polarization angle analysis performed by \cite{Fletcher:2004}

Key features of the magnetic field structure in M31 include a well-ordered large-scale magnetic field which is roughly aligned with the spiral arms of the galaxy \citep{Beck1996GalacticMR, Beck:2019}. The magnetic field in M31 is primarily concentrated in its disc \citep{Grave:1981}. Unlike some other galaxies that have a significant magnetic field in their halos, no halo field was detected in M31. The ordered magnetic field has a strength of around $5\mkG$ in the spiral arms. The total magnetic field strength, including both the regular and turbulent components, is around 6--$8\mkG$ \citep{Fletcher:2004, Beck:2019}. It is interesting to note that the central region of M31 ($r<0.5\kpc$) has an inclination and a position angle different from those of the disc and exhibits a low star formation rate. The regular magnetic field in the central region of M31 has the opposite direction compared to that in the disc \citep{2014A&A...571A..61G}. However, in this paper, we focus our analysis solely on the outer four rings (6--14\,kpc) of the M31 galaxy.

In this work, we apply the mode analysis technique of \cite{Fletcher:2004} to the latest radio polarization data of M31 from \cite{BBGM20} and $20.46$~cm wavelength data from \cite{1982A&A...106..121B}. Instead of the $\chi^2$ minimization used in earlier similar studies to determine the azimuthal Fourier-mode composition of the magnetic field, we use the Bayesian inference technique which is more appropriate
for highly non-linear models like the one used. We sample the joint posterior distribution of the Fourier mode parameters to study the correlation between the parameters of interest and reliably estimate the uncertainties. 

The paper is structured as follows. In \cref{sec:data}, we present the multi-wavelength radio polarization observations of M31 used for the analysis. \cref{sec:method} explains how we model the regular magnetic field in terms of Fourier modes in azimuth using the observed polarization angles. The Bayesian model selection approach and its validation is discussed in \cref{sec:BI}. The application of this method to the data is outlined in \cref{sec:implementation}. We present and discuss our results in \cref{sec:results}. Finally, we summarize our findings and provide conclusions in \cref{sec:summary}.

\begin{table*}
    \centering
    \caption{Wavelength, frequency, bandwidth, rms noise of polarized intensity and angular resolution {of the} data used for polarization and depolarization analysis. For the $20.46$~cm, the specifications for the VLA observation are listed. In the depolarization analysis (last row), we use $3.0\arcmin$ 
    for depolarization between $20.46$ and $6.18\cm$, and $5.0\arcmin$ for depolarization between $11.33$ and $6.18\cm$.}
    \begin{tabular}{lcccc}
        \toprule
        Central wavelength [cm] & $3.59$ & $6.18$ & $11.33$ & $20.46$ \\
        Central frequency [GHz] & $8.35$ & $4.85$ & $2.65$ & $1.46$ \\
        Bandwidth [MHz] & $1100$ & $500$ & $80$ & $50$ \\
        rms noise (PI) [mJy/beam] & $0.06\text{--}0.12$ & $0.05$ &$0.4$ &$0.025\text{--}0.05$ \\
        Resolution (HPBW) of the final maps & $1.4\arcmin$ & $2.6\arcmin$ & $4.4\arcmin$ & $45\arcsec$ \\
        Resolution (HPBW) for the polarization angle modelling & $3.0\arcmin$ & $3.0\arcmin$ & $4.4\arcmin$ & $45\arcsec$ \\
        Resolution (HPBW) for the depolarization analysis & $3.0\arcmin$ & $3.0\arcmin,\ 5.0\arcmin$ & $5.0\arcmin$ & $3.0\arcmin$ \\
        \bottomrule
    \end{tabular}
    \label{table:1}
\end{table*}


\section{Observations of M31} {\label{sec:data}}
The radio continuum data of M31 used in this study were obtained at four wavelengths: 3.59, 6.18, 11.33, and 20.46\,cm. The data at 3.59, 6.18, and 11.33\,cm were published by \citet{BBGM20}, while the 20.46\,cm data were originally presented by \citet{BBH98}. At 3.59\,cm (8.35\,GHz) and 11.33\,cm (2.645\,GHz), observations were conducted using the secondary-focus system of the Effelsberg 100-m radio telescope. The corresponding bandwidths were 1100\,MHz and 80\,MHz, respectively. Observations at 6.18\,cm (4.85\,GHz) were carried out using the secondary-focus system of the same telescope, with a bandwidth of 500\,MHz. These Effelsberg observations span the period from 2001 to 2012 and have a lower noise levels compared to earlier M31 data used by \citet{Fletcher:2004}, owing to recent advancements in radio technology.

\begin{figure*}
\centering
\includegraphics[clip, trim=5cm 2cm 6.2cm 2cm, width=0.75\linewidth] {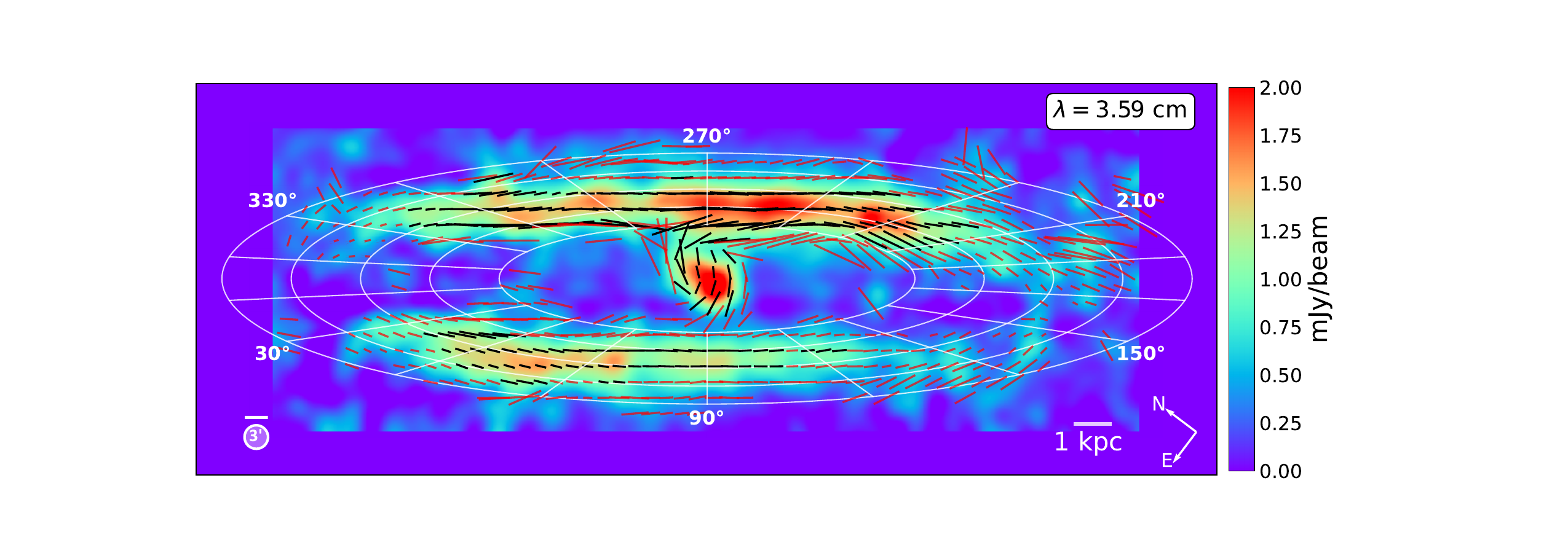}    
\includegraphics[clip, trim=5cm 2cm 6.2cm 2cm, width=0.75\linewidth] {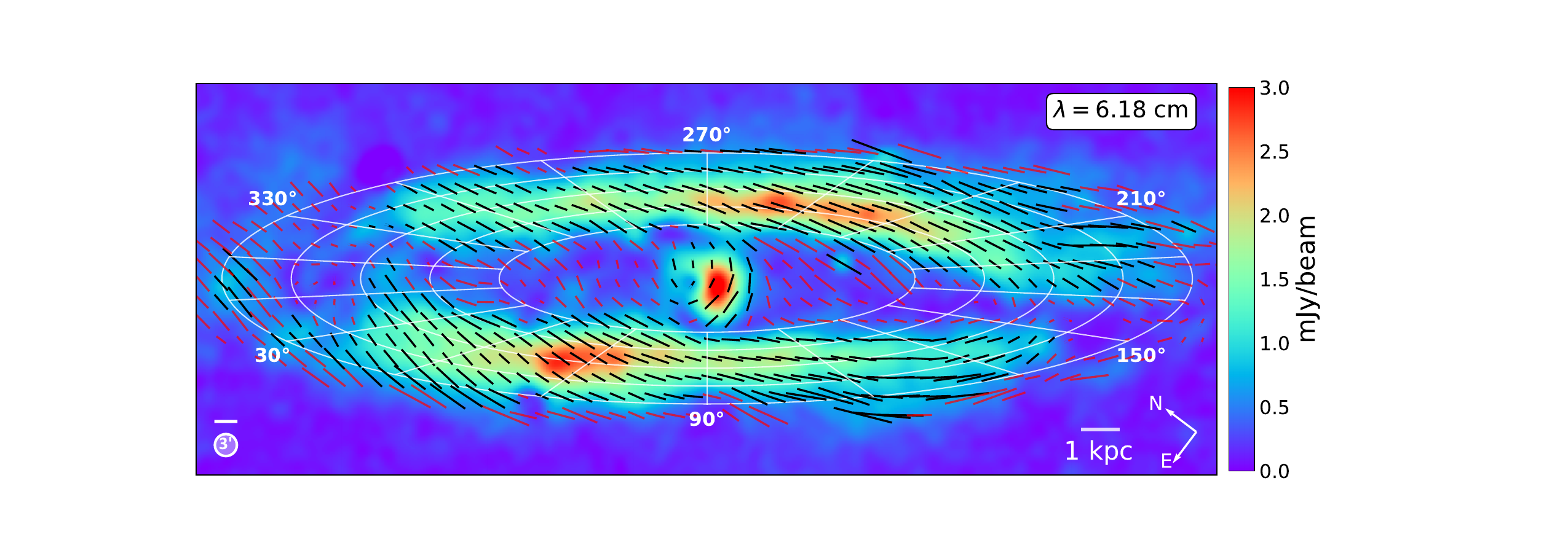}
\includegraphics[clip, trim=5cm 2cm 6.2cm 2cm, width=0.75\linewidth]{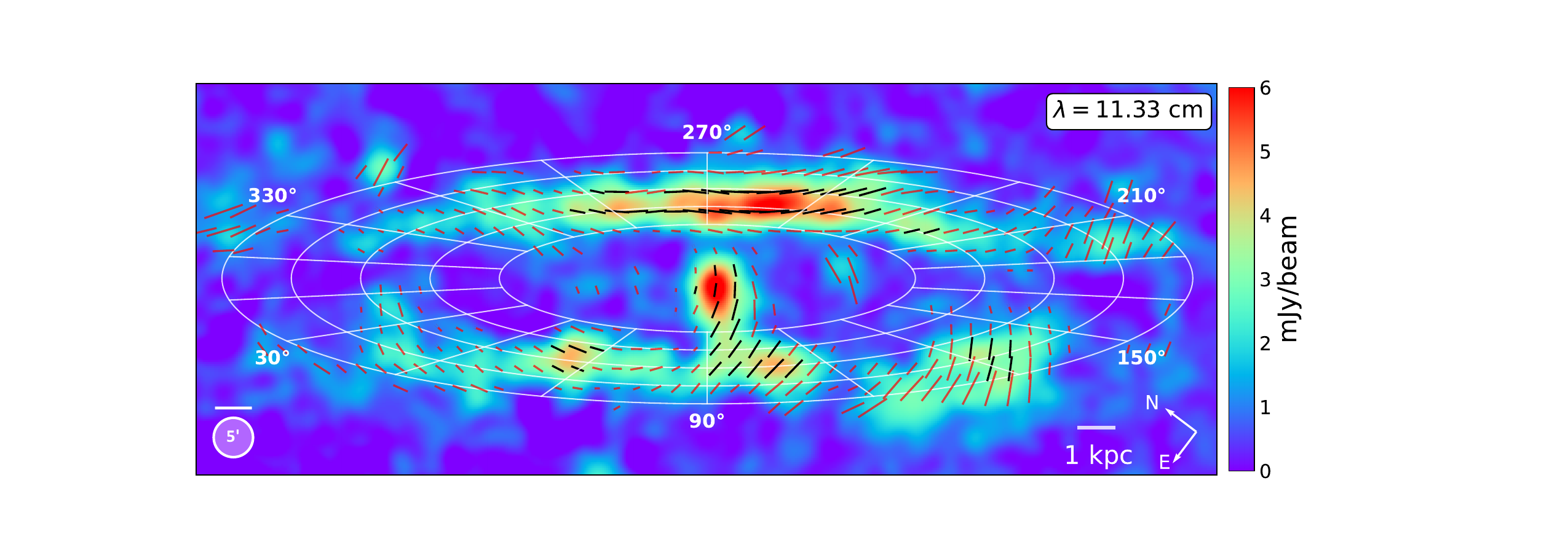}
\caption{Polarized intensity maps at various wavelengths in the sky plane. An overlay of azimuthal sectors of 20\deg\ width, as used in this paper, in the galactic plane is shown. The ellipses are the projections of circles in the galaxy plane, with radii of 6, 8, 10, 12, and 14\kpc. Polarized background sources have been subtracted. Straight line segments for the polarization orientation, with the length proportional to the degree of polarization, are shown only where total intensity is greater than three times the rms noise ($I \geq 3 \sigma_{I}$). They are shown in red where the polarized intensity is greater than three times the rms noise in the Stokes parameters $Q$ and $U$ ($\PI \geq 3 \sigma_{Q,U}$) and in black if the polarized intensity is greater than ten times rms noise in the Stokes parameters $Q$ and $U$. A segment of the length corresponding to 20 per cent polarization and the beam corresponding to the resolution used in this paper are shown at the bottom left, while the linear scale is at the bottom right.}
\label{fig:M31_data1}
\end{figure*}

In addition, we include L-band data at 20.46\,cm (1.46\,GHz) from \citet{BBH98}, which combines Effelsberg single-dish observations with interferometric data from the Very Large Array (VLA) in D configuration. The VLA observations, carried out between 1983 and 1992, have an angular resolution of $45\arcsec$. The final 20.46\,cm map includes large-scale total intensity structures from the Effelsberg data to compensate for the missing short baselines in the VLA data. The Effelsberg observations were performed at $1.4$~GHz (or 21.4~cm) with a bandwidth of 20~MHz between 1993 and 1995. A summary of the observational parameters, including central frequency, bandwidth, angular resolution (half-power beam width, HPBW), and root mean square (rms) noise, is provided in \cref{table:1}. The corresponding polarized intensity maps at $\lambda = 3.59$, 6.18, and 11.33\,cm are shown in \cref{fig:M31_data1}.

For the depolarization analysis in \cref{DeMe}, we use the derived non-thermal (synchrotron) intensity maps at three wavelengths from \cite{BBGM20}. The original map at $\lambda 3.59\cm$ was corrected for the contribution of anomalous microwave emission by multiplying the total non-thermal intensity by a factor of 0.831 \citep[see][for details]{Beck:2025}.

\subsection{Data averaging in rings and sectors}\label{sec:averaging}
The turbulent magnetic field strongly dominates at small scales \citep{Beck:2019}. However, the isotropic turbulent field is not observable in the polarized emission. Since we are interested in the large-scale structure of the magnetic field, we average the data over a sufficiently large area to eliminate the contribution of the random magnetic fields to the Faraday rotation. To do so, we split the galaxy into rings (annuli) of a fixed radial width of $2\kpc$, which is about 8.8\arcm\ along the major axis, in the galaxy plane and smooth the data over 18 sectors of equal angular size along the azimuthal coordinate $\theta$, after transforming to the cylindrical coordinate system $(r,\theta,z)$, where $z$ is the distance from the galactic mid-plane.
Sectors of $20\deg$ in width are wide enough to derive the standard deviation of polarization angle in each sector (with at least five data points per sector). There are typically five data points per beam diameter while each sector contains 2.4--4.5 beam areas at $8\text{--}14\kpc$ for the $3\arcmin$ beam width (1.1--2.1 for the $4.4\arcmin$ beam). Thus, individual angle measurements used in averaging over a sector are not completely statistically independent but this does not 
bias the Fourier mode analysis. Nevertheless, we avoid using narrower sectors (see, however, Appendix~\ref{SW}). The maximum number of azimuthal modes which can be fitted using 20\deg-wide sectors is discussed in Section~\ref{sec:method}. Using wider sectors would reduce the number of usable polarization angle measurements and cause stronger contamination by systematic gradients. The choice of a 20\deg\ sector size is also motivated by the need to maintain a sufficiently high signal-to-noise ratio given the available data.
The signal-to-noise ratio of polarized emission in the inner 1--$6\kpc$ is low, hence this region is excluded from our analysis. The mid-plane of the resulting cylindrical grid is shown in Fig.~\ref{fig:M31_data1}. 

From the polarimetric observations, we have the maps of the Stokes parameters $I$, $Q$, and $U$ at each wavelength. To compare them with the model (discussed in Section~\ref{sec:method}), we convert them into the polarization angle $\psi$ and polarized intensity $\PI$, using
\begin{equation}
	\psi = \frac{1}{2} \arctan  \frac{\langle U \rangle}{\langle Q \rangle}\,, 
\end{equation} 
\begin{equation}
\PI = \sqrt{ \langle Q \rangle ^2 + \langle U \rangle ^2 - (1.2 \sigma_{Q,U})^2} , \label{eq:2.2}
\end{equation}
where angular brackets denote the average value over all the pixels within the sector and $\sigma_{Q,U}$ is the rms noise in the measurements of $Q$ or $U$, respectively. The {term} $(1.2\sigma_{Q,U})^2$ in equation~\eqref{eq:2.2} represents the correction of positive bias due to noise \citep{wardle1974linear}. The error in {the} polarization angle is computed as the standard deviation of the polarization angles within one sector, including only pixels whose signal-to-noise ratio of $Q$ and $U$ is greater than 3, whereas the error in the polarized intensity is estimated as the standard deviation between all pixels in the sector. The average polarization angle $\psi$ within a sector is measured in the sky plane from the local outward radial direction. An appropriate coordinate transformation is performed to translate between the sky-plane and galactic-plane coordinates using the galaxy inclination angle of $i=75\deg$ and the position angle of the major axis $37\deg$ east of north \citep{chemin2009h}.

\section{Model of the magnetic structure}\label{sec:method}

The polarization angle $\psi$ of the synchrotron emission as a function of the azimuthal angle $\theta$ and wavelength $\lambda$ in a  ring sufficiently narrow to neglect the radial variation of the parameters can be written as 
\begin{equation}
 \psi(\theta)= \psi_0(\theta)+\lambda^2 \, \RM (\theta) + \lambda^2 \, \RM_{\rm fg} \ , \label{eqn:model}
\end{equation}
where $\psi_0$ is the intrinsic polarization angle, $\RM$ is the intrinsic Faraday rotation measure produced in the galaxy, and $\RM_{\rm fg}$ is the foreground Faraday rotation measure produced in the medium between the galaxy and the observer. The azimuthal angle
$\theta$ is  measured counter-clockwise in the galaxy plane from the northern major axis of the galaxy.
The quantities $\psi_0$ and $\RM$ depend on the regular magnetic field $\vec{B}$, inclination angle $i$ of the galaxy, the number density of the thermal electrons $n\el$, and the line-of-sight path length $L$ through the thermal disc in a given ring. $\RM_{\rm fg}$ is assumed to be independent of $\theta$ in a given ring. The validity of this assumption is discussed in Section~\ref{sec:summary}.

\label{anis_page}
The anisotropy of the random magnetic field, which also contributes to the polarized synchrotron emission, affects the inferred spiral pitch angle of the ordered magnetic field.
\citet[][e.g., their Table~1]{Beck:2025} find that the anisotropic random magnetic field in M31 is stronger than the mean (regular) magnetic field.
The average spiral pitch angles $p$ (defined in equation~\eqref{eqn:3.1}) of the ordered field
(dominated by the anisotropic field component) of $-(30^\circ\text{--}27^\circ)$ at $7<r<12\kpc$ \citep[Table~6 in][]{BBGM20}, measured from $\psi$, differ from the average pitch angles of $-(4^\circ\text{--}9^\circ)$ of the predominant axisymmetric mean magnetic field in the same radial range, measured from RM.
Therefore, the changes in $\psi_0$ due to the contribution of the anisotropic random magnetic field are significant in this radial range.
Making use of $\psi$ and RM simultaneously, the pitch angles of the predominant axisymmetric mean magnetic field at $6<r<12\kpc$ of $-(15^\circ\text{--}16^\circ)$ (\cref{tab:params_mcmc} below)
are in between the values from the two methods applied by \citet{BBGM20} and represent averages for both field components. The magnetic pitch angle at $12<r<14\kpc$ is slightly larger in magnitude but the anisotropy of the random field produced by the differential rotation \citep[section 4.4.1 of][]{SS21} is expected to be weaker in the outer galaxy.

The formalism used here is introduced and discussed in detail by \citet{Sokolov:1992} and \citet{BHKNPSS97}, who present it for a magnetic field containing all three cylindrical components. Here, we introduce it briefly, assuming that the vertical magnetic field component $B_z$ is negligible \citep{Fletcher:2004}. The remaining cylindrical components of $\vec{B} = (B_r, B_\theta)$ in the disc plane are expanded into the Fourier modes in $\theta$,
\begin{equation}\label{eqn:3.1}
\begin{split}
B_r &= \sum\limits_{m=0}^{N}B_m\, \sin{p_m}\, \cos[m(\theta-\beta_m)]\,,\\
B_\theta &=  \sum\limits_{m=0}^{N} B_m\, \cos{p_m}\, \cos[m(\theta-\beta_m)]\,,  
\end{split}
\end{equation}
where $B_m$ is the amplitude of the Fourier mode of the wave number $m$, $p_m$ is its spiral pitch angle (the angle between the field direction and the tangent to the circumference), $\beta_m$ is the azimuth where a non-axisymmetric mode is maximum, and  $N$ is the maximum number of modes required to fit the given data following the Nyquist--Shannon sampling theorem. For a horizontal magnetic field, $\psi_0$ is related to the model parameters as \citep[equation~A3 of][]{BHKNPSS97}
\begin{equation}
\psi_0 = \arctan \left[ \frac{\sin{(2\theta-p)}\sin^2{i}-(1+\cos^2{i})\sin{p}}{2\cos{p}\cos{i}}\right], \label{eq:3.2}
\end{equation}
where $p$ is the pitch angle of the total horizontal component of the regular magnetic field defined as $p= \arctan(B_r/B_\theta)$ with $-90^\circ<p\le 90^\circ$ and obtained using equation~\eqref{eqn:3.1}.
A trailing spiral has a negative pitch angle.

In the analysis of Faraday rotation, the amplitudes of the Fourier modes $B_m$ are convenient to characterise in terms of the related quantities
\begin{equation}\label{eq:3.3}
R_m = 0.81 \left(\frac{B_m}{1\mkG}\right) \left(\frac{\langle n_\e \rangle}{1\cm^{-3}}\right) \left(\frac{L} {1\p}\right) \frac{\text{rad}}{\text{m}^2}\,. 
\end{equation}
The parameters $B_m$ and $L$ can be understood as the equivalent amplitude of the magnetic mode and path length through the magneto-ionic layer.
Multiplying equation~\eqref{eqn:3.1} by $\langle n_\e\,\rangle L$, we obtain the corresponding quantities $R_r$ and $R_\theta$ which are the used to derive, via  the projection on the line of sight, the intrinsic Faraday rotation measure \citep[equation~A2 with $B_z=0$ of][]{BHKNPSS97}
\begin{equation}\label{Rpar}
R_\parallel =-(R_r\sin\theta+R_\theta\cos\theta)\sin i\,.
\end{equation}

Because of Faraday depolarization by magneto-ionic fluctuations, polarized radio emission at different wavelengths originates at different depths within the synchrotron emitting layer. For shorter wavelengths, Faraday depolarization is weak, making the medium transparent to polarized emission \citep{Fletcher:2004, BBGM20}. M31 does not have a detectable radio halo \citep{Grave:1981}, so $\RM$ likely depends on the magnetic field in the disc (large-scale halo magnetic fields may be present if the halo cosmic ray electron density is insufficient to produce detectable synchrotron emission, but we exclude halo magnetic fields from the model for simplicity). To allow for the fact that only a fraction of the synchrotron disc thickness is visible in polarized emission at a wavelength $\lambda$, we introduce the factor $\xi_\lambda$  \citep{BHKNPSS97} via 
\begin{equation}\label{eqn:3.4}
\RM = \frac{1}{2}\,\xi_\lambda \,R_\parallel\,,
\end{equation}
where $R_\parallel$ is the Faraday rotation measure accumulated through the whole disc thickness and the factor 1/2 gives the observable $\RM$ in a uniform slab or a magneto-ionic layer symmetric along the line of sight \citep{burn1966depolarization,sokoloff1998depolarization}.
This factor is different when the profile of magnetic field strength, thermal electron density, or density of cosmic-ray electrons in the magneto-ionic layer is not symmetric with respect to the galactic mid-plane \citep[sections 3.1--3.3 of][]{sokoloff1998depolarization}. As we argue in Section~\ref{DeMe}, the galaxy is not transparent to the polarized emission at $\lambda\lambda11.33$ and $20.46\cm$, so the symmetry is broken at these wavelengths. As a result, the relation between $\RM$ and $R_\parallel$ becomes significantly more complicated and wavelength-dependent. Therefore, the wavelength-dependent factor $\xi_\lambda$ allows not only for the reduced effective path length of the polarized emission but also for the effective asymmetry of the magneto-ionic layer. We assume that $\xi_3 = \xi_6 = 1$  at $\lambda\lambda3.59$ and $6.18\cm$ and $\xi_\lambda < 1 $ at the longer wavelengths (see Section~\ref{DeMe} for justification).

We fit the values of $\RM_\text{fg}$, $R_m$, $p_m$, and $\beta_m$, $0\leq m \leq N$, together with $\xi_{11}$ and $\xi_{20}$ to the observed polarization angles at all available wavelengths simultaneously. We adopt the value $L=4250\,$pc from \cite{Fletcher:2004} and $\langle n_\e\rangle=0.028, 0.032, 0.035$ and $0.021\cm^{-3}$ for the rings 6--8, 8--10, 10--12 and 12--14\kpc, respectively, from \citet{Beck:2025}, and use equation~\eqref{eq:3.3} to obtain the corresponding magnetic field amplitudes $B_m$. For an axisymmetric spiral (ASS) magnetic field ($N=0$), the model has three free parameters, $R_0$, $p_0$, and $\RM_{\rm fg}$ in addition to $\xi_{11}$ and $\xi_{20}$. For a combination of axisymmetric and bisymmetric field (BSS) fields ($N=1$), we have three additional parameters, $R_1$, $p_1$, and $\beta_1$. Each non-axisymmetric mode has three parameters, so the total number of fitted parameters in a model containing $n_m$ Fourier modes is $n=3n_m +2$, including $\xi_{11}$ and $\xi_{20}$.

The number of azimuthal modes which can be included into the model depends on the width of the sectors within which the data are averaged. 
To detect a periodic signal, the sampling rate must be twice the frequency of the signal according to the Nyquist--Shannon sampling theorem \citep{nyquist1928abridgment,shannon1949communication}. Therefore, we can detect no more than $n_m \leq 18/2 = 9$ Fourier modes. Since we omit some data points (see \cref{sec:exc_data_points}), there are effectively less than 18 points where the magnetic field is sampled. Therefore, we limit our analysis to  $n_m \leq 8$. The availability of the data at several wavelengths improves the reliability and accuracy of resulting magnetic field parameter estimates but this does not help to reveal higher  azimuthal modes as long as the sector width is not reduced.

The model for the polarization angles is highly non-linear and its parameter space is multi-dimensional. Therefore, the technique of residual minimisation used, e.g., by \citet{Sokolov:1992}, \citet{BHKNPSS97}, and  \citet{Fletcher:2004} is inefficient, depends on the initial parameter values used to initialize the minimisation, and does not allow for correlations between the parameters. We use the Bayesian inference technique for the fitting. 

\section{Bayesian inference}\label{sec:BI}

We introduce the notation $\alpha=(R_m, p_m, \beta_m, \RM_{\rm fg}, \xi_{11}, \xi_{20})$, $0\leq m\leq N$,  for the model parameters introduced in \cref{sec:method} and denote 
$\mathcal{M}$ the model itself and $\mathcal{D}$ the set of the polarization angle measurements to which it is fitted. The Bayes theorem expresses the posterior probability distribution $P(\alpha | \mathcal{D}, \mathcal{M})$ of the parameters given the data as
\begin{equation}
    P(\alpha | \mathcal{D}, \mathcal{M}) = \frac{\mathcal{L}(D| \alpha, \mathcal{M})\, \pi(\alpha | \mathcal{M})}{\mathcal{Z}(\mathcal{D} | \mathcal{M})} \ , \label{eq:4.1}
\end{equation}
where $\pi(\alpha | \mathcal{M})$ is the prior probability distribution of the parameters, $\mathcal{L}(D| \alpha, \mathcal{M})$ is the likelihood, and $\mathcal{Z}(\mathcal{D} | \mathcal{M})$ is the evidence. The evidence is defined as
\begin{equation}
    \mathcal{Z}(\mathcal{D} | \mathcal{M}) = \int \mathcal{L}(D| \alpha, \mathcal{M}) \, \pi(\alpha | \mathcal{M})\,  \dd \alpha \,. \label{eq:4.2}
\end{equation}
We assume a uniform prior for all the parameters within the allowed ranges:  
$|p_m| \le 90\deg$, and $|\beta_m| \le 180\deg/m$.

We adopt a Gaussian likelihood and assume that the data are sampled from a multivariate Gaussian distribution with a covariance matrix $\mathbf{\Sigma} (\alpha)$. The data uncertainties are known from the observations. 
When the path length through a turbulent magneto-ionic medium is much larger the turbulent scale, the polarization angle of radio emission affected by the Faraday rotation depends on $\RM$, the integral of the thermal electron density and magnetic field. Then the probability distributions of both the polarization angle and $\RM$ tend toward a Gaussian distribution (the central limit theorem) \citep{Kronberg:1994}.
The sampled polarization angle data can be written as $\boldsymbol{\psi}^\text{d} \sim \mathcal{N} (\boldsymbol{\mu} (\alpha), \mathbf{\Sigma} (\alpha))$ where $\mathcal{N}$ is the Gaussian distribution. For the likelihood of the model polarization angles $\boldsymbol{\psi}^\text{m}$ given by equation~\eqref{eqn:model} with $n=3n_m+2$  parameters, we adopt 
\begin{equation}
    \mathcal{L}_{\rm G} (\mathcal{D} | \alpha, \mathcal{M} ) = \frac{1} {(2 \pi)^{n/2}\sqrt{\mathbf{\Sigma}}} \, \text{exp} \left( - \frac{\chi^2}{2} \right ), \label{eq:4.3}
\end{equation}
where
\begin{equation}
 \chi^2 =  (\boldsymbol{\psi^\text{d}} - \boldsymbol{\psi^\text{m}})^T \mathbf{\Sigma}^{-1} (\boldsymbol{\psi^\text{d}} - \boldsymbol{\psi^\text{m}})\,. \label{eq:4.4}
\end{equation}
This choice is motivated by the fact that the observed polarization angles $\psi^\text{d}$ and, hence, the model residuals, $\boldsymbol{\psi^\text{d}} - \boldsymbol{\psi^\text{m}}$, have Gaussian probability distributions. We also reasonably assume that the observed polarization angle uncertainties $\sigma^\text{d}_\lambda(\theta_j)$ in different sectors $\theta_j$ are uncorrelated, so that the covariance matrix is a diagonal one. Then equation~\eqref{eq:4.4} reduces to
\begin{equation}
    \chi^2  = \mathlarger{\sum_{\lambda}} \mathlarger{\sum_{j}} \left [\frac{\psi^\text{d}_{\lambda} (\theta_j) -\psi^\text{m}_{\lambda} (\theta_j)} {\sigma^\text{d}_{\lambda} (\theta_j)} \right]^2, \label{eq:4.5}
\end{equation}
where the summations extend over the wavelengths and sectors within a given ring. 

We sample the posterior distribution given by equation~\eqref{eq:4.1} to obtain the joint probability distribution of the model parameters. We use the publicly available \textsc{cobaya} package written in \textsc{python} \citep{Torrado:2019} to sample the likelihood and derive the posterior distribution of the model parameters. \textsc{cobaya} uses the Metropolis--Hastings algorithm of the Markov Chain Monte Carlo (MCMC) method as described in \citet{Lewis:2013}. The first 30 per cent of the chains are discarded (`burn-in'). The algorithm stops when the Gelman--Rubin convergence parameter $\mathcal{R}$ becomes sufficiently close to unity or the total number of MCMC samples specified ($1\times10^6$) is attained, and we use $\mathcal{R} - 1 < 0.01$. As discussed below, the M31 data are fitted with $n_m\leq4$ azimuthal modes; for them, the MCMC converges before reaching the maximum number of samples. In order to validate the fitting procedure in \cref{sec:sim} we consider models with larger $n_m$, and for some of them the chain stops because of reaching the maximum number of samples.

We use the Bayesian complexity parameter $\mathcal{C}$ \citep{Spiegelhalter:2002} to 
determine the maximum number of the Fourier modes that the observed data can constrain \cite[see equation~(6) of][]{Kunz:2006}. In the case of a Gaussian likelihood, the complexity parameter is given by
\begin{equation}
     \mathcal{C} = \left <\chi^2(\alpha) \right >-\chi^2(\hat{\alpha}) \ , \label{eq:3.9}
\end{equation}
where the angular brackets denote the mean of $\chi^2$ taken over the posterior probability distribution and $\hat{\alpha}$ is the vector of the best-fitting parameters for a given model $\mathcal{M}$ which provide a maximum of the likelihood $\mathcal{L}_G$. We use the complexity parameter to decide whether or not a more complicated model is justified even if it provides a higher likelihood. Thus defined, the complexity parameter gives the effective number of degrees of freedom in the strongly non-linear models considered here.

When comparing two models, $\mathcal{M}_A$ and $\mathcal{M}_B$, we use both the complexity parameter and the model likelihood. Let $n_A$ and $n_B$ be the numbers of free parameters of the models $\mathcal{M}_A$ and $\mathcal{M}_B$, respectively, and suppose that $n_B < n_A$. The criteria for choosing the best model given the data are the following  \citep{Kunz:2006, Trotta:2008}: \textbf{(i)}~If $\log{\mathcal{L}_{\rm G}}(\mathcal{M}_A)\gg \log{\mathcal{L}_{\rm G}}(\mathcal{M}_B)$, model $\mathcal{M}_A$ is favoured over model $\mathcal{M}_B$. \textbf{(ii)}~If $\log{\mathcal{L}_{\rm G}}(\mathcal{M}_A)\approx \log{\mathcal{L}_{\rm G}}(\mathcal{M}_B)$ and $\mathcal{C}(\mathcal{M}_A)> \mathcal{C}(\mathcal{M}_B)$, the additional parameters of the model  $\mathcal{M}_A$ do not sufficiently improve the fit to the data, and we prefer model $\mathcal{M}_B$ as it has fewer parameters. \textbf{(iii)}~If $\log{\mathcal{L}_{\rm G}}(\mathcal{M}_A)\approx \log{\mathcal{L}_{\rm G}}(\mathcal{M}_B)$ and $\mathcal{C}(\mathcal{M}_A) \approx \mathcal{C}(\mathcal{M}_B)$,  both models are equally valid, and we need additional information to discriminate between them.

\begin{table}
\caption{Comparison between the input parameter values and the fitted ones for the example discussed in \cref{sec:sim}. The best-fitting parameters are shown together with their 
$1\sigma$ uncertainties.}
    \label{tab:simu_parameter}
    \centering
    \begin{tabular}{ccrr}
    \toprule
    Parameter & Unit & Input value & \text{Fitted value} \\
    \midrule
        \multicolumn{4}{c}{\bf Case I}  \\
         \bottomrule
          $\xi_{11}$ & & $0.8$ & $0.8\pm 0.1$\\
          $\xi_{20}$ & & $0.6$ & $0.6\pm 0.1$\\
          $\RM_\mathrm{fg}$ &$\mathrm{rad\,m^{-2}}$& $-130$ & $-127 \pm 2$ \\
          $R_0$ & $\mathrm{rad\,m^{-2}}$& 150 & $164 \pm 16$  \\
          $p_0$ & deg   & $-10$ & $-9 \pm 1$  \\
          \bottomrule
          \multicolumn{4}{c}{\bf Case II}  \\
        \bottomrule
          $\xi_{11}$ & & $0.8$ & $0.7\pm 0.1$\\
          $\xi_{20}$ & & $0.6$ & $0.5\pm 0.1$\\
          $\RM_\mathrm{fg}$ &$\mathrm{rad\,m^{-2}}$ & $-130$ & $-125 \pm 4$ \\
          $R_0$ & $\mathrm{rad\,m^{-2}}$& 150 & $190 \pm 29$  \\
          $p_0$ & deg   & $-10$ & $-8 \pm 3$ \\
          $R_1$ & $\mathrm{rad\,m^{-2}}$&100 & $124\pm 28$\\
          $p_1$ & deg   & 15 & $11 \pm 10$ \\
          $\beta_1$ & deg   & 20 & $5 \pm 11$ \\
          $R_2$ & $\mathrm{rad\,m^{-2}}$& 70 & $57 \pm 22$ \\
          $p_2$ & deg   &  $-30$ & $-45 \pm 16$ \\
          $\beta_2$& deg    &  $30$ & $16 \pm 9$ \\
          $R_3$ & $\mathrm{rad\,m^{-2}}$&  40 & $33 \pm 19$ \\
          $p_3$ & deg   & 10 & $19 \pm 44$\\
          $\beta_3$ &deg    &  $-20$ & $-8\pm 23$ \\
         \bottomrule
    \end{tabular}
\end{table}
\subsection{Validation of the inference procedure}\label{sec:sim}

\begin{figure}
    \centering
    \includegraphics[width=8cm]{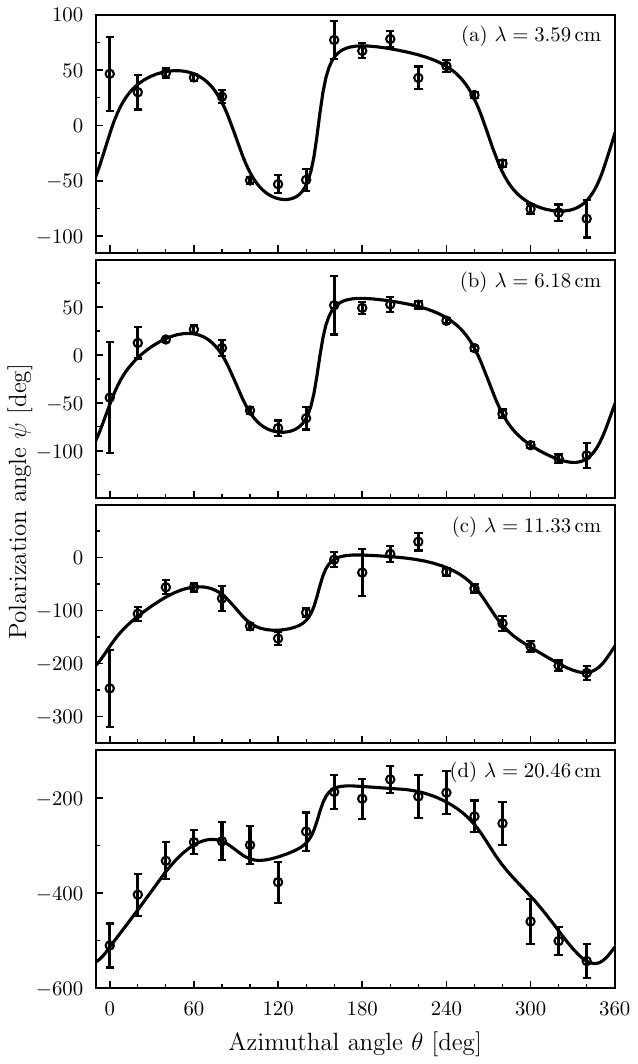}
    \caption{The simulated polarization angles for Case~II of \cref{sec:sim} (circles with error bars for the $1\sigma$ uncertainty) as a function of the azimuthal angle $\theta$ at four wavelengths, $\lambda \lambda 3.59$, $6.18$, $11.33$, and $20.46\cm$ from top to bottom. The black solid line represents the best-fitting model derived using the Bayesian inference.}
    \label{fig:4.1}
\end{figure}
 
We verify the Bayesian inference methodology using simulated polarization angles with realistic magnetic field model parameters and the uncertainties in the polarization angles from the actual M31 data. We generate samples of the polarization angles at the same four wavelengths and the same sectors as the data adopting a regular magnetic field represented by a number of azimuthal Fourier modes. We consider two magnetic field structures: Case~I, an axisymmetric spiral (ASS) magnetic field, $m=0$ and Case~II, a superposition of azimuthal modes up to $m=3$ with parameters shown in \cref{tab:simu_parameter}. The Faraday depolarization factors used to generate the synthetic data are $\xi_{3}=1$ {and} $\xi_{6}=1$ while $\xi_{11}$ and $\xi_{20}$ are presented in Table~\ref{tab:simu_parameter}. We have verified that the conclusions of this section are not sensitive to the exact choice of $\xi_{\lambda}$. We add a realistic scatter into the simulated polarization angles using the observed $1\sigma$ errors in Ring 3 (10--$12\kpc$) and assuming 
Gaussian statistics for the error.
In both cases, we discard the first 30 per cent of the MCMC chains and use the convergence criterion $\mathcal{R} - 1 < 0.01$. The simulated polarization angles are fitted using models with a varying number of azimuthal modes $N$. We first consider the simplest model which only has the $m=0$ mode and add one Fourier mode at a time until the criteria formulated above are satisfied.

\begin{figure}
\centering
\includegraphics[width=\linewidth]{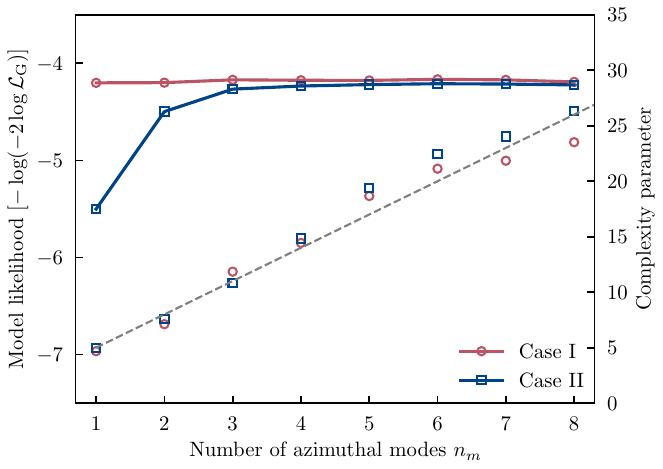} 
\caption{The variation of the maximum likelihood $\mathcal{L}_\text{G}$ (left-hand axis, continuous lines; higher likelihood corresponds to a larger value of the quantity shown) and the complexity parameter $\mathcal{C}$ (right-hand axis, open symbols) with the number of magnetic modes $n_m$ in the validation procedure of \cref{sec:sim}: Case I (red curve and circles) and Case~II (blue curve and squares). 
The dashed grey line corresponds to  $\mathcal{C} =3n_m+2$, the number of the model parameters.}
\label{fig:4.2}
\end{figure}

\paragraph*{Case I.}
We draw 12,800 MCMC samples of the polarization angles to derive the fitted parameters given in \cref{tab:simu_parameter}, which match the input values very well. 

\paragraph*{Case II.}
To accommodate the increased number of parameters, we draw 153,440 MCMC samples to obtain results given in \cref{tab:simu_parameter} and shown in Fig.\ \ref{fig:4.1}. The agreement between the input and best-fitting parameters is, understandably, not as good as in the case of a simpler magnetic field structure in Case~I but remains quite satisfactory. 

 Figure~\ref{fig:4.2} shows the model likelihood and complexity parameter as functions of the number of the Fourier modes included in the Bayesian inference. In Case~II, the complexity parameter grows almost linearly as $\mathcal{C} \approx 3n_m+2$, whereas the likelihood increases rapidly with $n_m$ up to $n_m=4$, where $3n_m$ is the number of parameters in a model containing $n_m$ azimuthal modes. The likelihood slowly decreases for $n_m>1$ in Case~I. 
 The complexity parameter of Case~I deviates from $3n_m+2$ already for $n>2$. Using the criteria of Bayesian model selection introduced in \cref{sec:method}, we can conclude that the simulated polarization angle data prefer $n_m=1$ (a model including the $m=0$ mode alone) for Case~I and $n_m=4$, i.e., $m=0,1,2,3$ for Case~II. When the model becomes too complicated ($n_m$ increases too far), the complexity parameter deviates from an approximately linear growth with $n_m$ and $\mathcal{C}<3n_m+2$.

 We perform this analysis for the other rings and various values of $\xi_\lambda$ to arrive at the same conclusions. In summary, we have verified that, under the assumption of a Gaussian likelihood, we can constrain the number of azimuthal modes $n_m$ for the two cases considered in our analysis using the simulated polarization angle data. Our simulation setup does not include all the complications present in the real data, such as the presence of local features, additional RM contribution, etc.
 
\begin{figure*}
\centering
\includegraphics[width=\linewidth]{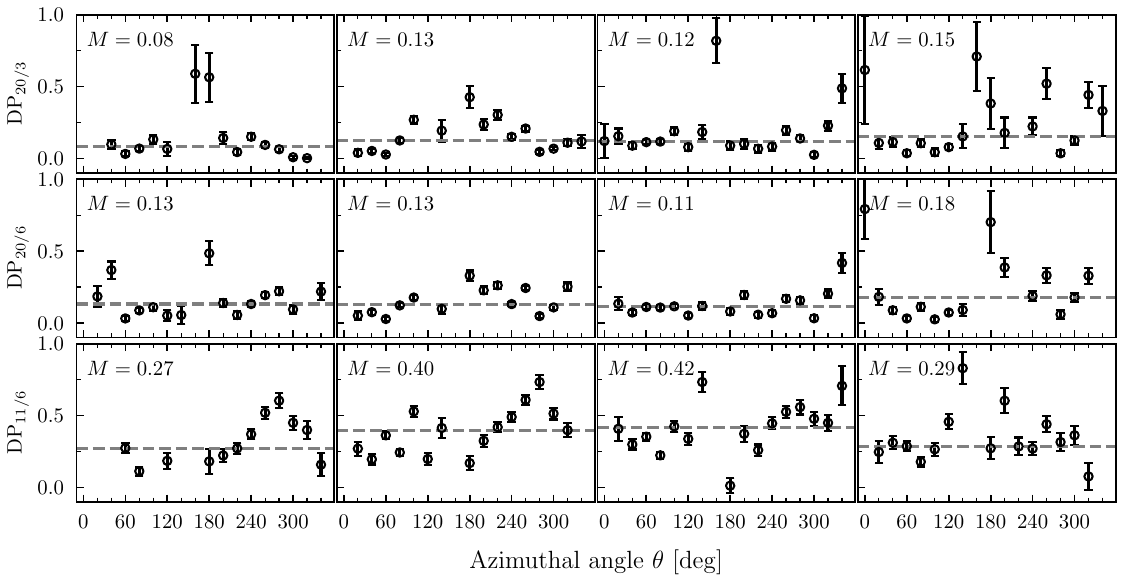}
\caption{The variation of depolarization as a function of the azimuthal angle: $\DP_{20/3}$ at 3\arcm\ HPBW (top panel),  $\DP_{20/6}$ at 3\arcm\ HPBW (middle panel), and $\DP_{11/6}$ at 5\arcm\ HPBW (bottom panel). The median depolarization across the azimuthal angles per radial range is shown with the grey dashed line. Left to right: the radial ranges 6--8, 8--10, 10--12 and 12--14\kpc.    }
\label{fig:sigma_rm}
\end{figure*}

\section{Implementation}\label{sec:implementation}

Faraday depolarization of the radio emission of M31 is significant at the longer wavelengths used in our analysis, especially at $\lambda20.46\cm$. As a result, the polarization signal at this wavelength is accumulated at a fraction $\xi_\lambda$ of the total path length through the disc of M31, significantly shorter than the path lengths at $\lambda 3.59, \lambda 6.18$, and $\lambda 11.33\cm$. Therefore, we first analyse the polarized intensity to verify that the depolarization at the shorter wavelengths is sufficiently weak to be negligible. In our analysis, the depolarization factors $\xi_{11}$  and $\xi_{20}$ are fitted together with the other model parameters of the mean magnetic field using the Bayesian inference approach described in \cref{sec:BI}.

\subsection{Depolarization}\label{DeMe}

\citet{Fletcher:2004} attributed the depolarization pattern obtained from the observations available to them to a rotation measure gradient in a Faraday screen, as suggested by a periodic variation of the depolarization with $\theta$ (see their figures~8 and 9). The more sensitive observations at $\lambda\lambda 3.59, 6.18,$ and $11.33\cm$ used here do not support that interpretation. In particular, there are only weak (if any) signs of the periodicity, which makes Faraday depolarization by random fields (which do not vary with $\theta$ as strongly as $\RM$) more plausible. The depolarization is defined as the ratio of the degrees of polarization $p(\lambda)$ at two wavelengths,
\begin{equation}\label{DePo}
\DP_{\lambda_1/\lambda_2} = \frac{p(\lambda_1)}{p(\lambda_2)}\,.
\end{equation}
When the depolarization is caused by the Faraday rotation of either mean or random magnetic field, $\DP_{\lambda_1/\lambda_2}<1$ when $\lambda_1>\lambda_2$, and $\DP_{\lambda_1/\lambda_2}$ is smaller when the ratio $\lambda_1/\lambda_2$ is larger. The stronger is depolarization, the smaller is $\DP$ when $\lambda_1>\lambda_2$. 

Figure~\ref{fig:sigma_rm} shows the depolarization between $\lambda \lambda20.46$ and $3.59\cm$ at 3\arcm\ HPBW, between $\lambda \lambda 20.46$ and $6.18\cm$  at 3\arcm\ HPBW and between $\lambda \lambda 11.33$ and $6.18\cm$  at 5\arcm\ HPBW. If the depolarization was caused by the gradient of $\RM$ due to the mean magnetic field, $\DP$ would be larger near the major axis, $\theta=0\deg,180\deg$, where $\RM$ changes rapidly with $\theta$ because of the large inclination of M31 to the line of sight \citep{Fletcher:2004}. Some values of $\DP$ near $\theta=180\deg$ in Fig.\ \ref{fig:sigma_rm} are relatively large but this does not happen near $\theta=0\deg$.
Data points which deviate strongly from their typical values have larger uncertainties.

The median depolarization values ($M$) in the rings 6--8, 8--10, 10--12, and 12--$14\kpc$ are $0.08$, $0.13$, $0.12$, and $0.15$ for $\DP_{20/3}$, $0.13$, $0.13$, $0.11$, and $0.18$ for $\DP_{20/6}$, and $0.27$, $0.40$, $0.42$, and $0.29$ for $\DP_{11/6}$, respectively. The difference between $\DP_{20/6}$ and $\DP_{20/3}$ is not large and given that the Faraday depolarization at $\lambda3.59\cm$ is negligible, we neglect it at $\lambda6.18\cm$ as well, thus adopting $\xi_3=\xi_6=1$. Indeed, the degrees of polarization at $\lambda\lambda3.59$ and $6.18\cm$ are very similar \citep[section~2.5 and fig.~4 of][]{Beck:2025} and depolarization at these wavelengths is due to the wavelength-independent effects (the scatter in $\psi_0$ due to the random magnetic field within the telescope beam) rather than Faraday rotation. Meanwhile, \citet{Beck:2025} show that the degrees of polarization at $\lambda\lambda 11.33$ and $20.46\cm$ are significantly lower, apparently because of the Faraday depolarization.

Since the depolarization does not vary much with $\theta$, we conclude that the magneto-ionic fluctuations (Faraday dispersion in the synchrotron-emitting region and screen) are the main sources of depolarization. The origin of weaker depolarization near the major axis visible in Fig.\ \ref{fig:sigma_rm} remains to be understood.

The depolarization mechanism at $\lambda\lambda 11.33$ and $20.46\cm$ can also involve significant depolarization by the regular (mean) magnetic field and thus be not only stronger but also more complicated than at the shorter wavelengths. Therefore, we treat $\xi_{11}$ and $\xi_{20}$ as independent parameters to be fitted. 

\subsection{Data filtering}\label{sec:exc_data_points}

Although the magnetic field model includes high-order modes and, thus, can tolerate rapid changes of $\psi$ with $\theta$, a few polarization angle measurements (15 out of 288, specified in \cref{tab:outliers}) cannot be fitted. These data points only occur at the three shorter wavelengths and are mostly located near the major axis, $\theta=0\deg$ and $180\deg$, where the polarization angle changes rapidly with $\theta$. Because of the high inclination of M31 to the line of sight, the partial overlap of telescope beams affects the observational results more strongly near the major axis, leading to an underestimated uncertainty of the observed polarization angle. This does not happen at $\lambda20.46\cm$ where the resolution is significantly higher than at the shorter wavelengths (see \cref{table:1}). In the ring 6--8\kpc, the data at  $\theta=280\deg\text{--}40\deg$ at $\lambda3.59\cm$ are unreliable due to technical issues in the subtraction of the foreground radio emission. An exception is the sector at $\theta=100\deg$, $r=6\text{--}8\kpc$ located near the minor axis. This sector exhibits a significantly higher polarization than its neighbours (Fig.\ \ref{fig:M31_data1}). This suggests that this data point is affected by a strong local fluctuation.

The data from the sectors shown in \cref{tab:outliers} are treated as outliers and excluded from any further modelling of the polarization angles.

\begin{table}
    \centering
    \caption{Data points excluded from the analysis as outliers.}
    \label{tab:outliers}
    \begin{tabular}{cccc}
    \toprule
Radial & \multicolumn{3}{c}{Azimuthal angle $\theta$ [deg]} \\
range [kpc]  &$\lambda3.59\cm$    &$\lambda6.18\cm$ &$\lambda11.33\cm$\\
      \midrule
        6--8    &0, 20, 140, 340   &0        &100, 140, 160\\
        8--10   &0                 &0, 340   &--\\
        10--12  &--                &--       &160\\
        12--14  &--                &340 &0, 160\\
        \bottomrule
    \end{tabular}
\end{table}

\begin{figure}
    \centering
    \includegraphics[width=\columnwidth]{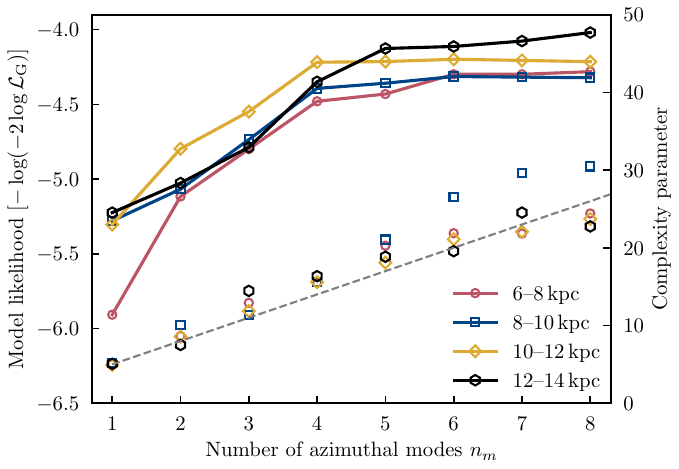}
    \caption{The model likelihood (curves with symbols) and complexity parameter (symbols) as functions of the number of Fourier modes for $r=$6--8\,kpc (red circles), 8--10\,kpc (blue squares), 10--12\,kpc (orange diamonds) and 12--$14\kpc$ (black hexagons). 
    The complexity parameter increases almost linearly, as $\mathcal{C}=3n_m+2$ up to $n_m=5$. Together with the variation of the likelihood, this suggests $n_m=3$ for $r=6\text{--}12\kpc$ and $n_m=4$ for $r=12\text{--}14\kpc$.}
    \label{fig:complexity_plot}
\end{figure}

\begin{figure*}
    \centering
    \includegraphics[width=\textwidth]{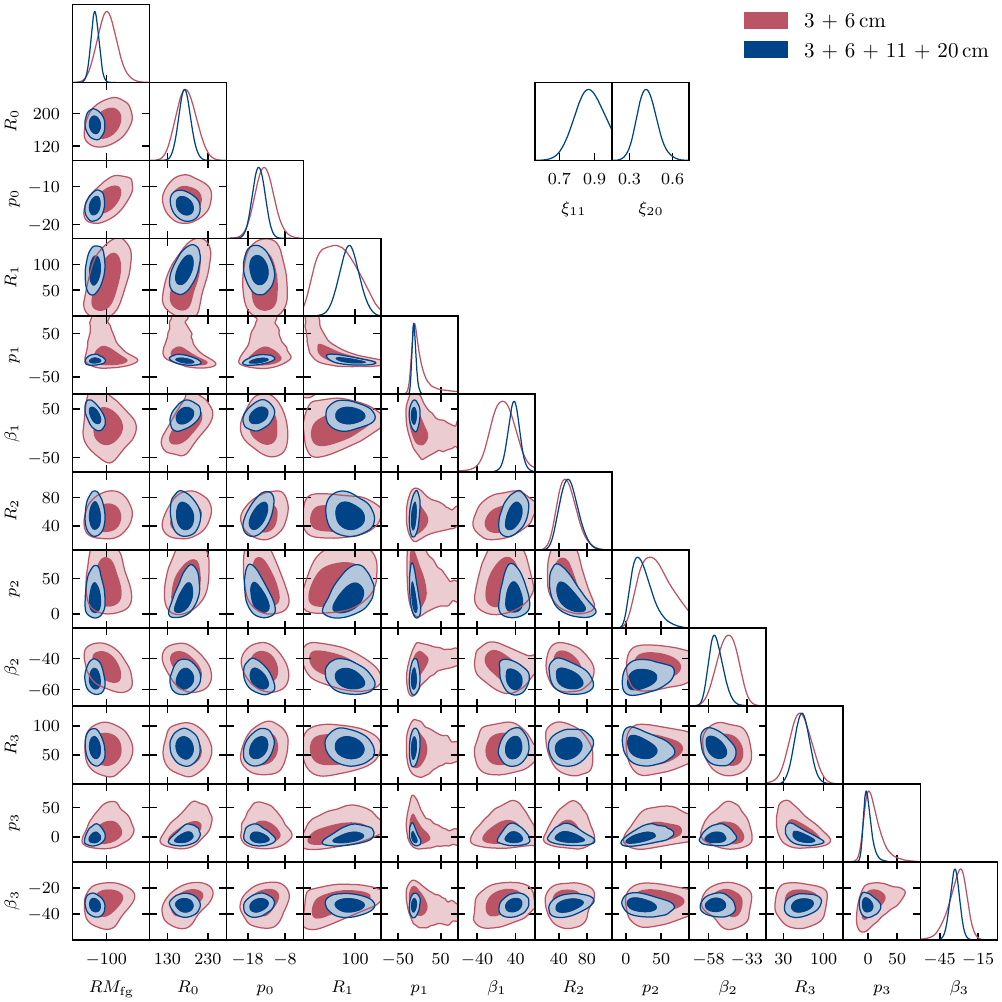}
    \caption{The likelihood variation with the model parameters for the radial ring 8--10\,kpc. The shape of the likelihood contours indicates that there is no significant correlation between the parameters. The $1\sigma$ and $2\sigma$ regions are indicated by the dark and light blue colours for the fit involving all four wavelengths and dark and light red for the fit relying on $\lambda\lambda3.59$ and $6.18\cm$ data alone (the latter are independent of $\xi_{11}$ and $\xi_{20}$ -- see Appendix~\ref{apdx1} for details). The marginal probability distributions of each parameter are shown on the right (in the corresponding colours). The likelihood has a single maximum in all cases confirming the convergence to the model parameters shown in \cref{tab:params_mcmc}.}
    \label{fig:mcmc_triangular_2}
\end{figure*}

\begin{table*} 
\centering 
\caption{The mean values and $1\sigma$ uncertainties of the model parameters $\xi_{11}$, $\xi_{20}$, $\RM_\text{fg}$, $R_m$, $p_m$ and $\beta_m$ obtained by fitting the four wavelengths together, and the thermal electron number density $\langle n_\e \rangle$ derived by \citet{Beck:2025}.} 
\begin{tabular}{ccccccc}  
\toprule 
Azimuthal mode &Parameter& Unit & \multicolumn{4}{c}{Radial range [kpc]}\\ 
 \cline{4-7}
number & & & \multicolumn{1}{c}{6--8} & \multicolumn{1}{c}{8--10} & \multicolumn{1}{c}{10--12} & \multicolumn{1}{c}{12--14} \\ 
\midrule
  \addlinespace[2pt]
&  $\langle n_\e \rangle$ & cm$^{-3}$ & 0.028 & 0.032 & 0.035 & 0.021 \\
  \addlinespace[2pt]
  & $\xi_{11}$ &  & $0.8\pm 0.1$ & $0.9\pm 0.1$ & $ 0.9\pm 0.1$  & $ 0.8 \pm 0.1$ \\
& $\xi_{20}$ &  & $0.2\pm 0.1$ & $0.4\pm 0.1$ & $0.6\pm 0.1$ & $0.7 \pm 0.1$ \\
    &$\RM_\text{fg}$ &rad\,m$^{-2}$ &$-134\pm4$  &$-108\pm3$  &$-90\pm4$  &$-101\pm5$\\ 
\midrule 
\multirow{2}{*}{$m=0$}  &$R_0$&rad\,m$^{-2}$ &$157\pm20$  &$173\pm20$  &$217\pm20$  &$199\pm23$\\ 
&$p_0$  &deg &$-16\pm2$  &$-15\pm2$  &$-15\pm2$  &$-22\pm2$\\ 
\midrule 
\multirow{3}{*}{$m=1$}  &$R_1$  &rad\,m$^{-2}$ &$57\pm20$  &$89\pm20$  &$104\pm17$  &$49\pm19$\\ 
&$p_1$  &deg &$-28\pm10$  &$-12\pm5$  &$30\pm10$  &$22\pm20$\\ 
&$\beta_1$  &deg &$69\pm30$  &$36\pm10$  &$32\pm6$  &$20\pm 14$\\ 
\midrule 
\multirow{3}{*}{$m=2$}&$R_2$&rad\,m$^{-2}$ &$55\pm10$  &$55\pm10$  &$69\pm20$  &$107\pm21$\\ 
&$p_2$  &deg &$38\pm16$  &$24\pm16$  &$-49\pm9$  &$-34\pm7$\\ 
&$\beta_2$  &deg &$-48\pm4$  &$-53\pm5$  &$42\pm6$  &$21\pm6$\\ 
\midrule 
\multirow{3}{*}{$m=3$}  &$R_3$  &rad\,m$^{-2}$ &$111\pm18$  &$63\pm13$  &$60\pm10$  &$72\pm14$\\ 
&$p_3$  &deg &$7\pm6$  &$0\pm8$  &$9\pm11$  &$13\pm12$\\ 
&$\beta_3$  &deg &$-23\pm2$  &$-33\pm4$  &$-36\pm4$  &$-50\pm3$\\
\midrule 
\multirow{3}{*}{$m=4$}  &$R_4$  &rad\,m$^{-2}$ &--  &--  &--  &$36\pm12$\\ 
&$p_4$  &deg &--  &-- &--  &$38\pm27$\\ 
&$\beta_4$  &deg &--  &--  &--  &$-43\pm4$\\
\midrule 
Derived parameter & $B$ & $\!\mkG$ & 1.7 & 1.7 & 1.7 & 2.7 \\
\bottomrule 
\end{tabular} 
\label{tab:params_mcmc}
\end{table*} 
\begin{figure*}
\includegraphics[width=\textwidth]{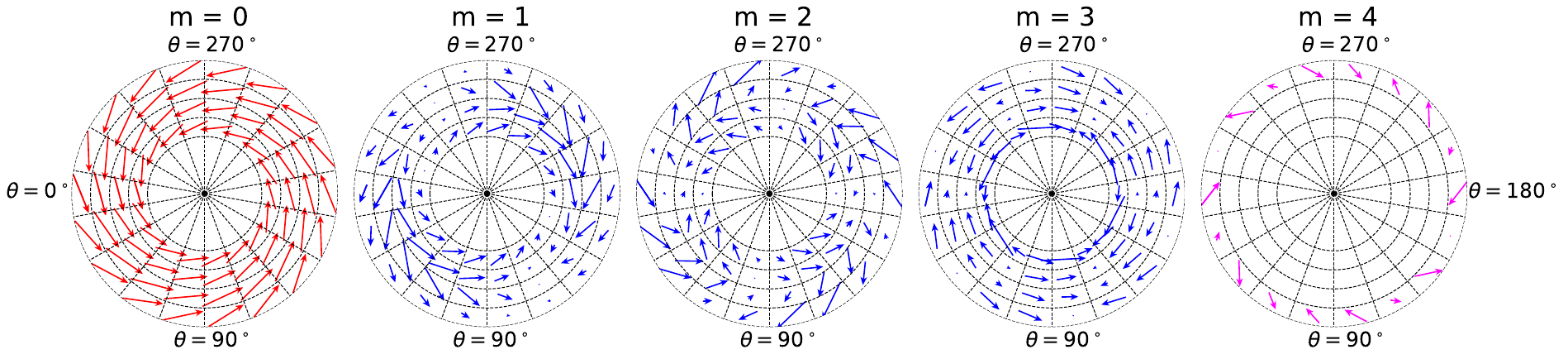}
\caption{The magnetic field structure of the individual modes $m=0$, 1, 2 and 3, from left to right with the vector length proportional to $R_m\cos[m(\theta-\beta_m)]$. The lengths of the vectors shown in red ($m=0$) are downscaled by a factor of four, and those in blue ($m=1,2,3$) are downscaled by a factor of two with respect to the ones in pink.}
\label{fig_model_3611_1}
\end{figure*}

\begin{figure*}
\centering
\includegraphics[width=\linewidth]{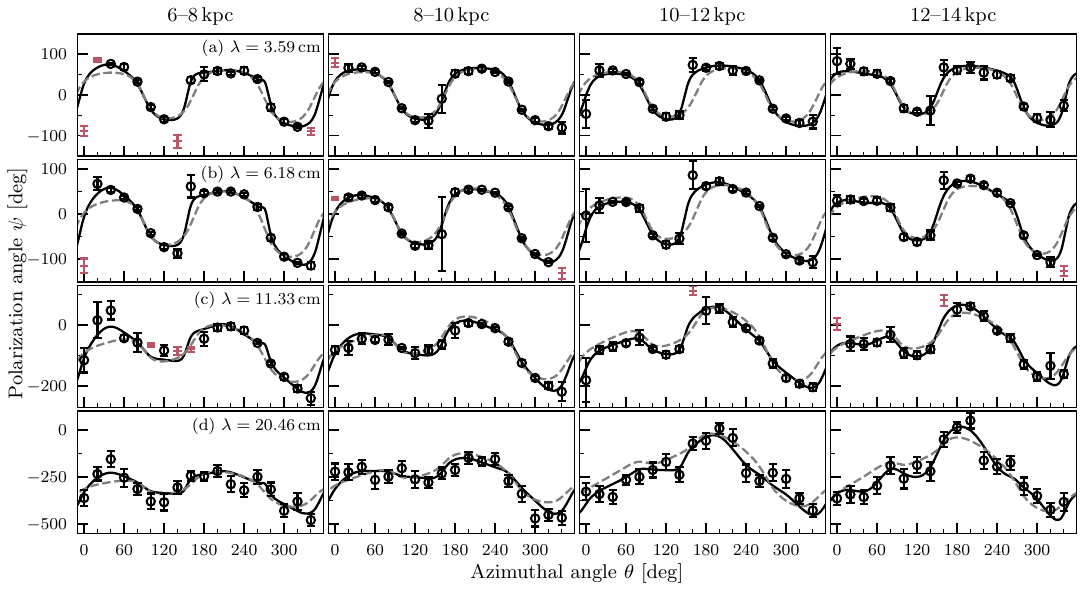}
\caption{The observed (symbols with $1\sigma$ error bars) and fitted (solid) polarization angles versus the azimuthal angle for $r=6\text{--}8$, 8--10, 10--12, and 10--$12\kpc$ (left to right) and $\lambda=3.59$, 6.18, 11.33, and $20.46\cm$ (top to bottom). The best-fitting mode combinations have $n_m=4$ for $r=6\text{--}12\kpc$ and $n_m=5$ for $r=12\text{--}14\kpc$. The grey dashed line represents the contribution of dominant $m=0$ Fourier mode alone. The excluded data points are shown in red. The $\pm180\deg$ uncertainty in the observed $\psi$ has been used to obtain a continuous variation of $\psi$ with $\theta$. }
\label{fig:model}
\end{figure*}

\section{Results}\label{sec:results}

We applied the MCMC analysis to the observed M31 polarization angles at four wavelengths for the radial ranges 6--8, 8--10, 10--12, and 12--14\kpc, respectively). Fig.\ \ref{fig:complexity_plot} shows the variation of the complexity parameter and the likelihood with the number of modes included in the model. At 6--12\kpc, subdominant contributions from the modes $m=1,2,3$ are required to fit the measured polarization angles. One more Fourier mode $m=4$ is required in the radial range $r=12\text{--}14\kpc$. For $n_m\leq5$, the convergence condition $\mathcal{R} - 1 < 0.01$ is satisfied for all four rings confirming that the MCMC chains are converged to a reasonable accuracy. Models containing other possible mode combinations are discussed in \cref{appendix1}. 

To verify that the results are not too sensitive to the value of $\xi_{20}$ (the only one among $\xi_\lambda$ which differs significantly from unity), we repeated the analysis excluding the data at $\lambda20.46\cm$. This fitted parameter is difficult to estimate from any physical arguments and, thus, to confirm that its fitted value is consistent with any other information available. The results obtained using only the data at $\lambda\lambda3.59$ and $6.18\cm$ are discussed in \cref{apdx1}. The larger is the number of the wavelengths used, the smaller are the uncertainties of the resulting parameter values but the results obtained using various wavelength combinations agree within $1\sigma$ errors. 

Figure~\ref{fig:mcmc_triangular_2} shows, for the 8--$10\kpc$ ring, the contours of the model likelihood in the planes defined by all possible parameter pairs, as well as the marginal probability distribution for each parameter (the picture is similar in the other rings as shown in Appendix~\ref{apdx1}). The likelihood contours are shown for both the model based on all four wavelengths (blue) and on the shorter wavelengths $\lambda\lambda3.59$ and $6.18\cm$ (red). The fits based on the shorter wavelengths do not involve $\xi_{20}$. The similarity of both the likelihood contours and the marginal probability distributions confirm that the results are not very sensitive to the value of $\xi_{20}$.  The likelihood has an isolated maximum at the fitted parameter values in all cases confirming the convergence of the MCMC chains to the solution. The shape of the contours suggests that there are no significant cross-correlations between the model parameters.

\cref{tab:params_mcmc} presents the mean values and $1\sigma$ uncertainties of the model parameters obtained from the fit to the data at all four wavelengths, and 
Fig.\ \ref{fig_model_3611_1} shows the magnetic field structure in each azimuthal model while Fig.\ \ref{fig:model} shows the resulting dependence of $\psi$ on $\theta$ where the dashed grey line shows the contribution of the axisymmetric mode ($m=0$) alone. As shown in  Fig.~\ref{fig_res_all}, the difference  $\psi^\text{d}-\psi^\text{m}$ between the observed and fitted polarization angles is mostly less than the $1\sigma$ uncertainty of the polarization angle measurements $\sigma_\psi$: the rms 
values of $(\psi^\text{d}-\psi^\text{m})/\sigma_\psi$ are 1.10, 0.99, 0.87 and 0.81 for the rings 6--8, 8--10, 10--12 and 12--14\kpc, respectively.  The difference of the polarization angles varies with $\theta$ and $r$ randomly, without any systematic trends. The axisymmetric part of the magnetic field is responsible for the main features in the distribution of the polarization angle along azimuth and the magnetic modes with $m>0$ are required only because of local, relatively rapid variations of $\psi$ with $\theta$.

The foreground Faraday rotation measure $\RM_\text{fg}$ is accumulated in the Milky Way. The estimated value of the Milky Way's RM contribution towards M31 from the full-sky rotation measure map is $-83\pm 21\radm$ \citep{Hutschenreuter:2022}. At around $-100\radm$ for the outer three rings, it is broadly consistent with earlier estimates \citep{ruzmaikin1985magnetic,han1998new,berkhuijsen2003polarized,BBGM20} but the value for the innermost ring (6--8\kpc) is noticeably different, $\RM_\text{fg}=-134\pm4\radm$.
It cannot be excluded that $\RM_\text{fg}$ varies due to local magneto-ionic fluctuations in the Milky Way, which are known to occur on angular scales smaller than the size of M31 \citep{Pandhi2025}. The H$\alpha$ map of \citet{Ogle2025} shows extended patches in front of M31 with fluctuations on various spatial scales. If this ionized gas carries magnetic fields, it would contribute to $\RM_\text{fg}$.

\begin{figure*}
    \centering
    \includegraphics[width=0.7\textwidth]{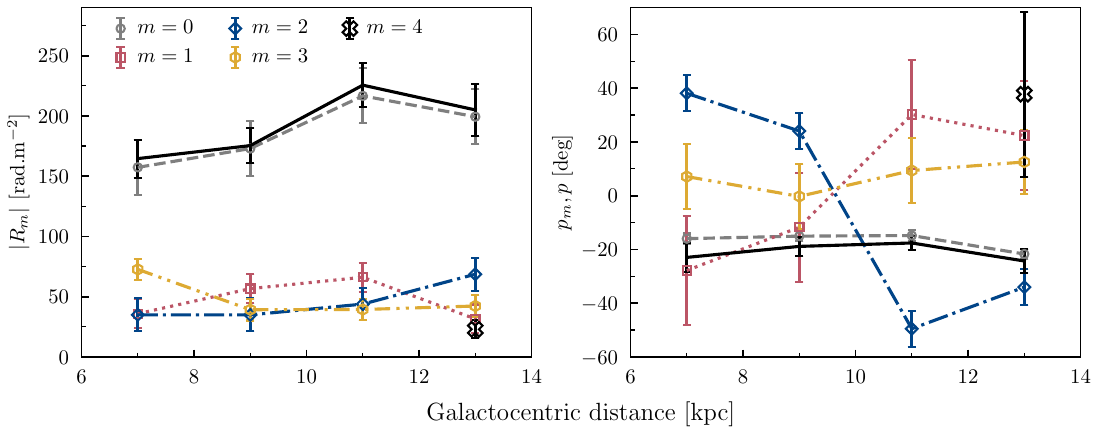}
    \caption{The variation with the galactocentric radius of the amplitudes of the individual magnetic modes $|R_m|$ and their pitch angles $p_m$ ($m=0,1,2,3,4$) from \cref{tab:params_mcmc} together with the total amplitude $|R| \propto{(B_r^2+B_{\theta}^2)^{1/2}}$ and the pitch angle of the total regular magnetic field $p=\arctan(B_r/B_{\theta})$ obtained from equation~\eqref{eqn:3.1} (written in terms of $R_m$):  $m=0$ (dashed/grey), $m=1$ (dotted/red), $m=2$ (dash-dotted/blue), $m=3$ (dash-double-dotted/yellow), $m=4$ (a single point in the outer ring, black) and the total amplitude $|R|$ (solid/black). $|R|$ and $p$ are functions of $\theta$; their azimuthal averages are shown.}
    \label{fig:var_params}
\end{figure*}

\begin{figure*}
    \centering
    \includegraphics[width=0.955\columnwidth]    {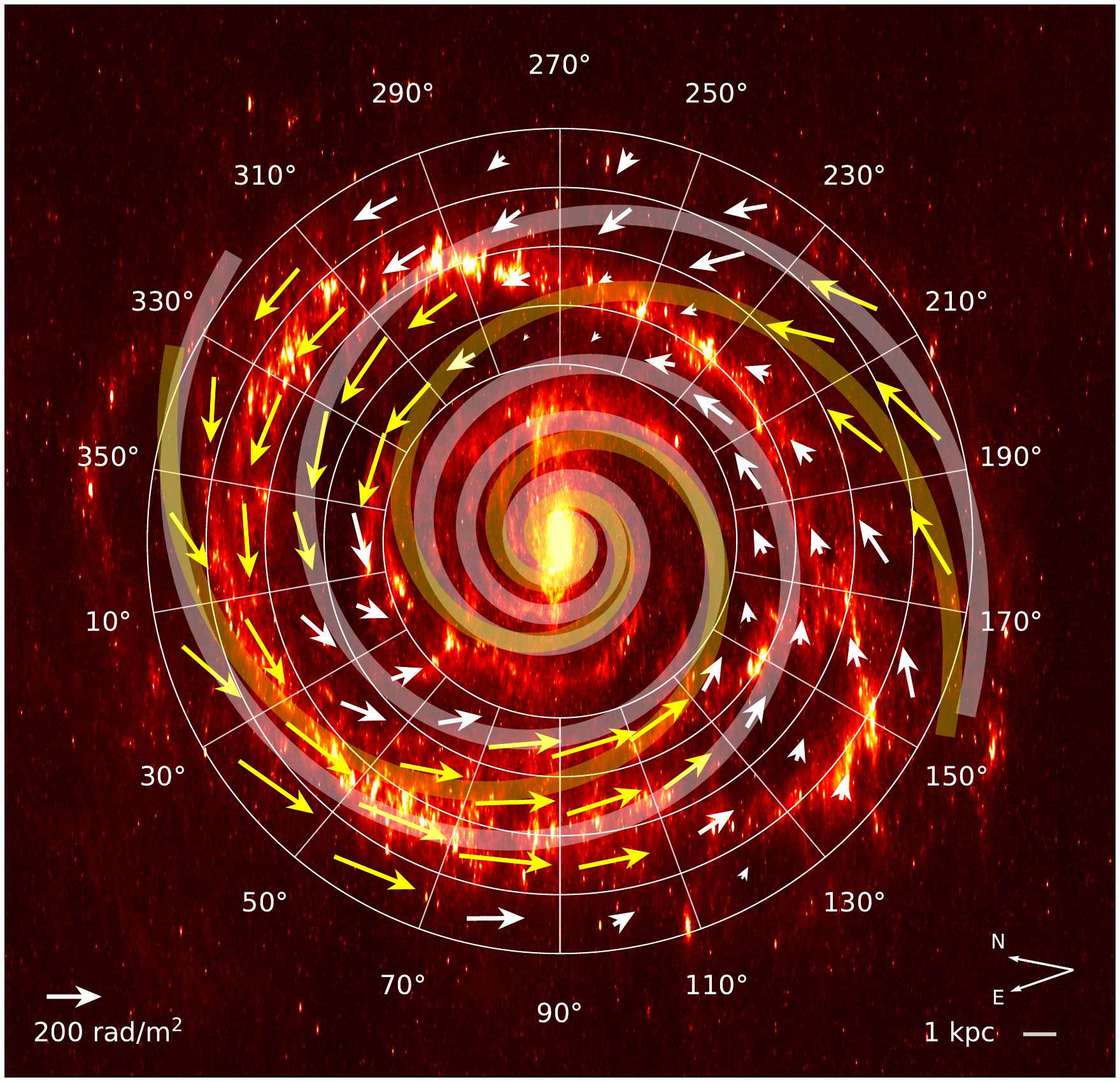}
    \includegraphics[width=0.94\columnwidth]{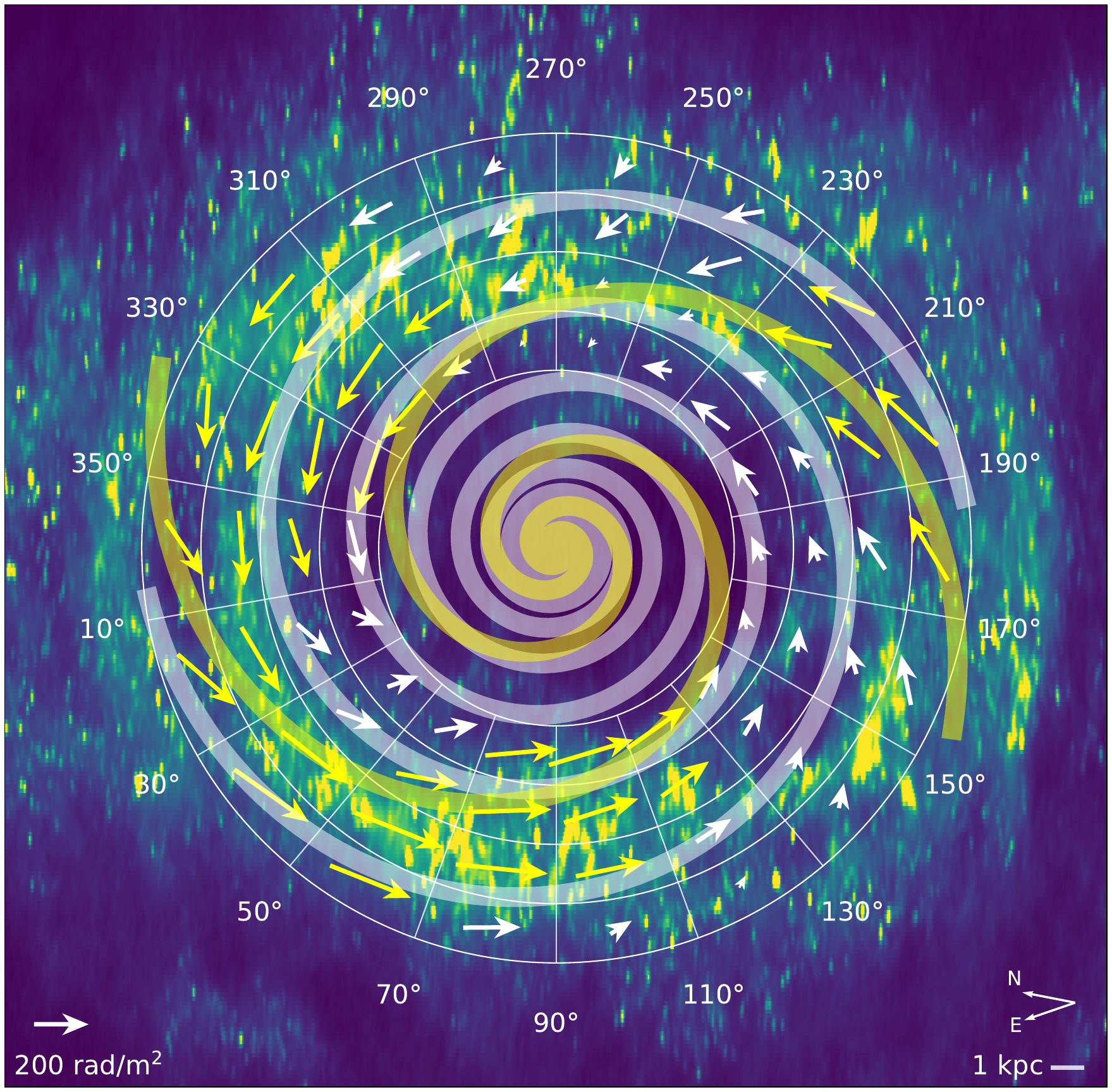}
    \caption{\textbf{Left:} the face-on view of the magnetic field vectors for the model of \cref{tab:params_mcmc} superimposed on the deprojected infrared ($\lambda24\,\upmu$m) image of M31 with the spiral arms identified by \citet{G+06} in grey with pitch angles of $-9\deg$ and $-9.5\deg$ and phase $166\deg$ (see the text for details). The image is oriented as in Fig.\ \ref{fig:M31_data1}.  
    The vector lengths are proportional to $R_m\cos m(\theta-\beta_m)$, and those which exceed 50 per cent of the maximum in each ring are shown in yellow. The polar grid used in the paper is in white. 
    The spirals shown in yellow have the pitch angle of $-16\deg$, close to that of the $m=0$ mode, thus representing $m=0$ magnetic lines, shown at the phase angles of 30\deg\ and 210\deg\ of the $m=1$ mode. 
    \textbf{Right:} similar to the image on the left but with the deprojected H\,\textsc{i} spectral line ($\lambda21$ cm) image of M31 as the background  \citep{braun2009}, with the spiral arms detected in neutral hydrogen shown in grey \citep[][fig.~7, with the fitted pitch angle $-6.7\deg$, shown at the phase angles 120\deg\ and 300\deg]{Braun91}.}
\label{fig:polar}
\end{figure*}

\subsection{Magnetic field structure and strength}

The axisymmetric magnetic field ($m=0$) dominates at all galactocentric distances, its amplitude $R_0$ is more than twice as large as that of the higher-order modes. Fig.\ \ref{fig:var_params} shows the variation of the mode amplitudes $|R_m|$ and their pitch angles $p_m$ with the galactocentric radius. 
Also shown is the azimuthally averaged strength of the total regular magnetic field $(B_r^2+B_\theta^2)^{1/2}$ (in terms of $R_m$ rather than $B_m$) and its pitch angle: both are close to the corresponding parameters of the $m=0$ mode.
Together with the total regular magnetic field, its axisymmetric part has a maximum at $r=10\text{--}12\kpc$ in the bright radio ring, but the amplitudes of the higher-order modes are significantly smaller and vary little with $r$. The pitch angles of both the total regular field and its axisymmetric part remain practically constant as $r$ varies, so the magnetic lines are trailing logarithmic spirals. The pitch angles of the higher modes vary in a rather disordered manner and can be both positive and negative. This suggests that the axisymmetric part of the magnetic field is produced by a systematic physical process (presumably, the mean-field dynamo action) but the higher-order modes emerge from a distortion of the axisymmetric magnetic field by local perturbations in the magneto-ionic medium such as the hole at $\theta \simeq 149\deg$, $r\simeq10\kpc$ resulting from the passage of M32 through the disc of M31 \citep{G+06}. The bisymmetric mode $m=1$ introduces some asymmetry in the magnetic structure, with the magnetic field slightly stronger around $\theta=\beta_1=20\text{--}50^\circ$ in all the rings. The mode $m=2$ might be associated with a two-armed spiral pattern but the lack of any systematic variation of its pitch angle $p_2$ and phase $\beta_2$ do not support such an interpretation.\label{m-dyn}

The amplitudes of the magnetic field strength in each azimuthal mode can be obtained from the values of $R_m$ ($m=0,1,2,\ldots$) shown in \cref{tab:params_mcmc} using equation~\eqref{eq:3.3}. For the thermal electron number densities $\langle n_\e\rangle$, we use the estimates of \cite{Beck:2025}, also given in Table~\ref{tab:params_mcmc}.
The scale height $h$ of the thermal disc of $2h\simeq1100$\,pc \citep{Fletcher:2004} yields the path length $L\simeq 2h/\cos i \simeq 4250\p$, and the relation between $R_m$ and the strength of the large-scale regular field $B_m$ of equation~\eqref{eq:3.3} becomes
\begin{equation}\label{BmRm}
 B_m =4\mkG \left(\frac{R_m}{100\radm}\right) \left(\frac{\langle n_\e\rangle}{0.03\cm^{-3}}\right)^{-1}  \left(\frac{L}{1\kpc}\right)^{-1}.
\end{equation}
The strength of the regular magnetic field of mode $m=0$ is the largest at about $2.7\mkG$ in the ring 12--$14\kpc$ and slightly weaker at about $1.7\mkG$ in the inner rings. These strengths are smaller than those estimated by \cite{Fletcher:2004}
who however derived the field strengths from polarized intensities which are dominated by anisotropic random fields \citep{Beck:2025}. The values of the regular magnetic field strength obtained here are consistent with the values obtained by \citet{Beck:2025} using Faraday rotation measures, too.

The magnetic field structure and its relation to the $24\,\upmu$m infrared emission (a tracer of interstellar dust) is illustrated in the left-hand panel of Fig.\ \ref{fig:polar}. \citet{G+06} identify two logarithmic trailing spiral arms with the pitch angles $-9\deg$ and $-9.5\deg$ which trace most of the dust structures at $r\lesssim 10\kpc$ shown in the figure in grey. The spiral structure in H\,\textsc{i} is less well pronounced but it
can be represented with a two-armed trailing pattern with the pitch angle $p=-6.7^\circ$ at  $6.2<r<18.6\kpc$ \citep{Braun91}. The relation between the regular magnetic field and H\,\textsc{i} is illustrated in the right-hand panel of Fig.\ \ref{fig:polar}.
The axisymmetric part of the magnetic field also represents a logarithmic spiral but but it is more open, with the pitch angle $p_0=-15\deg\pm2\deg$. As in other galaxies, the pitch angles of the magnetic field and spiral arms are not the same since they depend on different galactic parameters \citep[Section~13.10 of][and references therein]{SS21}. The regular magnetic field of M31 at the scales captured by our analysis shows little correlation with the spiral patterns visible in the dust and neutral gas distributions. As shown in the left-hand panel of Fig.\ \ref{fig:polar} magnetic field can be stronger (vectors shown in yellow) both in and between the spiral arms delineated by either the interstellar dust or neutral hydrogen.

The large-scale magnetic field is not uniform along the bright radio ring at $r\approx 11\kpc$ being stronger in the northern part of the galaxy. This asymmetry extends across the whole range of $r$ and is visible in the original maps shown in Fig.\ \ref{fig:M31_data1}. It causes the presence of the $m=1$ (bisymmetric) magnetic mode which has maximum magnitude at $\theta=\beta_1=20\text{--}40\deg$ and $\theta=\beta_1+180\deg=200\text{--}220\deg$ 
at $r=8\text{--}14\kpc$.

Although the dust and gas appear to be well mixed in M31 \citep{G+06}, the strength of the regular magnetic field is clearly not well correlated with the dust emission often being high where the $24\,\upmu$m emission is quite low. This is not surprising since, apart from the gas density, the strength of the regular magnetic field depends on the galactic angular velocity, large-scale velocity shear due to differential rotation, and turbulent velocity among other parameters \citep{CSS14}. Although the infrared \citep{G+06,F+12}, ultraviolet \citep{T+05} and CO \citep{N+06} emission are strongly concentrated in spiral arms suggesting a significant 
arm--interarm contrast in the gas density, the regular magnetic field remains predominantly axisymmetric. Of course, spiral arms add structure to the magnetic field  and the magnetic modes with $m>0$ are caused by these deviations from the axial symmetry.

\begin{figure*}
    \centering
    \begin{tabular}{cc}
    \includegraphics[width=0.35\textwidth]{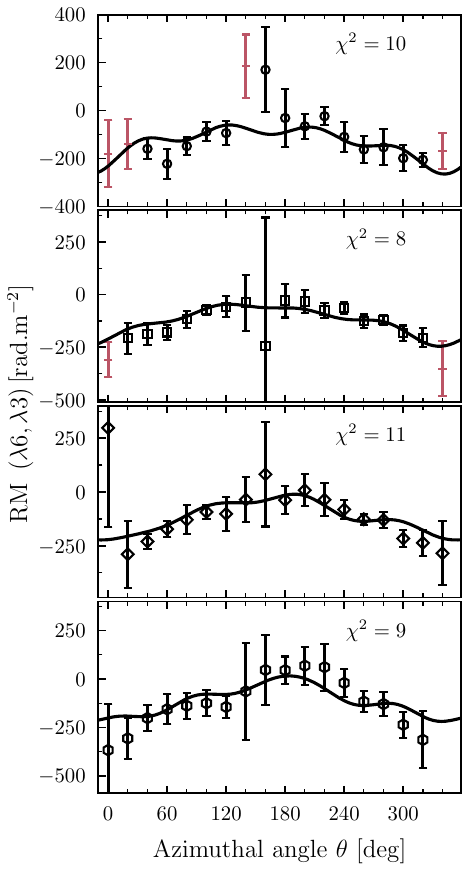}  &
   \includegraphics[width=0.35\textwidth]{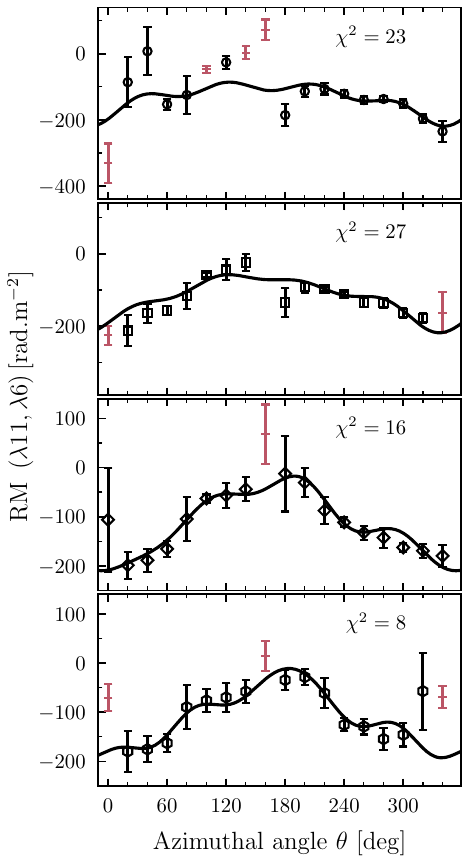}
   \end{tabular}
    \caption{The values of RM with $1\sigma$ error bars obtained from equation~\eqref{eqn:rm_ex} between $\lambda \lambda3.59$ and $6.18\cm$ at 3\arcmin\ resolution (left panel) and $\lambda \lambda 6.18$ and $11.33\cm$ (on the right) at 5\arcmin\ resolution,  for the radial ranges from 6--$8\kpc$ (top) to 12--$14\kpc$ (bottom). The black solid lines show     
    RM derived using the best-fitting parameters of \cref{tab:params_mcmc}. 
    The $\chi^2$ test confirms that the two distributions agree well if only the data points shown in red are excluded; they also excluded from the fitting of the polarization angles.}
    \label{fig:rm_36}
\end{figure*}

The difference in the path lengths in polarized emission at the shorter and longer wavelengths also  affects the Faraday rotation measure, and an independent test of the model is provided by the values of $\RM$ obtained from the observed polarization angles at wavelengths $\lambda_1$ and $\lambda_2$ as
\begin{equation}\label{eqn:rm_ex}
    \RM \approx \frac{\psi_{\lambda_1}-\psi_{\lambda_2}}{\lambda_1^2-\lambda_2^2}\,,
\end{equation}
which is an approximation since $\RM$ generally is not a linear function of $\lambda^2$ in an inhomogeneous and asymmetric magneto-ionic layer emitting synchrotron \citep{sokoloff1998depolarization}. In our case, the synchrotron disc is not visible through its whole depth in the polarized emission at the longer wavelengths (11.33\,cm and 20.46\,cm) due to the Faraday depolarization, creating an apparent asymmetry.
Figure~\ref{fig:rm_36} shows $\RM$ obtained from equation~\eqref{eqn:rm_ex}
together with that obtained using the best-fitting magnetic field parameters from \cref{tab:params_mcmc}, confirming the high quality of the fits. Furthermore, this confirms that the contribution of anisotropic random magnetic field to $\psi_0$ can be neglected because $\psi_0$ affects the magnetic field model obtained here but does not affect $\RM$.

\section{Discussion and conclusions}\label{sec:summary}

With the recently observed, more sensitive radio data at the wavelengths $3.59$, $6.18$, and $11.33\cm$, along with previously observed $\lambda20.46\cm$ polarization angle data, we revise and extend the Fourier mode analysis for the regular magnetic field between the galactocentric distances $r=6$ and $14\kpc$ using the Bayesian inference framework, an approach more robust and general than in earlier similar analyses.

We confirm earlier findings that the large-scale magnetic field in M31 is predominantly axisymmetric but reveal significant deviations from axial symmetry at all galactocentric distances explored. The axisymmetric part of the magnetic field is consistent with predictions of the mean-field dynamo theory \citep[][and references therein]{Nazareth+25}. The non-axisymmetric magnetic modes ($m>0$ for the azimuthal wave number) reflect local magnetic features at a kiloparsec scale resulting from the overall asymmetry of the galaxy, its spiral pattern and other perturbations, such as the hole at $r=8\text{--}12\kpc$, $\theta=130^\circ\text{--}150^\circ$ produced by the passage of M32 \citep{G+06}. Since the magnitudes and pitch angles of the non-axisymmetric modes vary rather irregularly with $r$ (\cref{tab:params_mcmc}), they do not appear to be directly supported by the dynamo action. The general picture is fully consistent with mean-field galactic dynamo models which include complicated spiral patterns and bars \citep[e.g.][]{MS91,SM93,Moss+07,Chamandy+13a,Chamandy+13b,Chamandy+14a,Moss+15,Chamandy+15}.

Synchrotron emission at $\lambda11.33\cm$ and especially $\lambda20.46\cm$ is depolarized by Faraday rotation in both the regular and random magnetic fields. Since such parameters of the magneto-ionic medium as the correlation scale of $\RM$, the number density of thermal electrons and correlation between the magnetic and electron density fluctuations are highly uncertain, it is difficult to estimate the amount of depolarization \textit{a priori}. Instead, we suggest to parametrize the depolarization using the parameters $\xi_{11}$ and $\xi_{20}$ to allow for the effective reduction of the path length at $\lambda\lambda11.33$ and $20.46\cm$. These parameters are fitted together with the magnetic field parameters. The results are quite satisfactory (Appendix~\ref{Dep}), demonstrating a modest amount of depolarization at $\lambda11.33\cm$ ($\xi_{11}\approx 1$) and significant depolarization at $\lambda20.46\cm$ ($\xi_{20}$ is significantly smaller than unity -- see \cref{tab:params_mcmc}).

Our conclusions are as follows: 
\begin{enumerate}
    \item We confirm that the large-scale (regular) magnetic field in M31 is dominated by the axisymmetric spiral (azimuthal wave number $m=0$) but detect higher-order modes $m \leq 4$. The $m=0$ mode amplitude varies smoothly with the galactocentric radius with a maximum at $r=10\text{--}12\kpc$, where the field strength is about $2.7\mkG$. It has a nearly constant pitch angle $p=-(15\deg\text{--}16\deg)$ (a trailing logarithmic spiral) at $r=6\text{--}12\kpc$ with a more open spiral $p=-22\deg$ at $r=12\text{--}14\kpc$. This dominant part of the magnetic field is consistent with the expectations of the galactic mean-field dynamo theory. The magnetic field extends to $r<6\kpc$ and $r>14\kpc$.
    
    \item The higher modes ($m>0$) have much smaller amplitudes than the $m=0$ mode which vary irregularly with $r$, have positive and negative pitch angles which also vary strongly with $r$. This suggests that they are not related to any large-scale dynamo action but rather arise from various deviations form axial symmetry in the galactic velocity field and gas density distribution. The $m=1$ (bisymmetric) mode is slightly stronger than those with $m>1$ and its presence reflects the fact that the magnetic field is stronger in the north-eastern part of the galaxy.   
        
    \item The amount of depolarization of the radio emission at $\lambda\lambda11.33$ and $20.46\cm$ varies little with the azimuthal angle in the galactic plane. With earlier observational data, the depolarization was attributed to the rotation measure $\RM$ gradients in a Faraday screen. The more sensitive data do not support this explanation. The lack of a strong, systematic azimuthal variation of the depolarization suggests that Faraday depolarization by random magnetic fields dominates, although some contribution from the large-scale field cannot be excluded. Both the depolarization and deviations of the galactic magneto-ionic layer from symmetry along the line of sight (especially significant at the longer wavelength where the galaxy is not transparent to the polarized radio emission) are parametrized with the factors $\xi_{11}$ and $\xi_{20}$ which are considered free parameters fitted together with the other parameters of the magnetic field model. Thus, our analysis is independent of the nature of the depolarization.
    
    \item The foreground Faraday rotation measure $\RM_\text{fg}$ attributable to the Milky Way is not consistent across all the radial rings.  At $r=8\text{--}14\kpc$, $\RM_\text{fg}\approx -100\radm$ but stronger foreground Faraday rotation with $\RM_\text{fg}\approx -134\radm$ apparently occurs at $r=6\text{--}8\kpc$, probably related to inhomogeneous magnetic-ionic gas in the foreground of the Milky Way.
\end{enumerate}

\section*{Acknowledgements}

We thank Dr Shabbir Shaikh for the useful discussion and help with the Bayesian inference code. This work was supported by the Science and Engineering Research Board, Department of Science and Technology, Government of India, grant number SERB/ECR/2018/000826. The computations in this paper were run on the Hercules cluster at NISER supported by Department of Atomic Energy of the Government of India. S.\ S.\ is supported by the National Postdoctoral Fellowship of the Science and Engineering Research Board (SERB), ANRF, Government of India (File No.: PDF/2023/000594). We thank the anonymous referee for careful reading of the manuscript and useful suggestions.

\section*{Data availability}
The software used in this work is publicly available: \textsc{Astropy} \citep{Astropy_2013A&A, astropy_2018}, \textsc{matplotlib} \citep{Matplotlib}, \textsc{cobaya} \citep{Torrado:2020dgo} and \textsc{GetDist} \citep{Lewis:2019}.
We use the publicly available M31 data (\url{https://cdsarc.cds.unistra.fr/ftp/J/A+A/633/A5/fits/}) for the analysis of this work. 

\bibliographystyle{mnras}
\bibliography{sed} 

\appendix

\section{Verification and variations of the magnetic field model}\label{apdx1}

\begin{figure*}
\centering
\includegraphics[width=\textwidth]{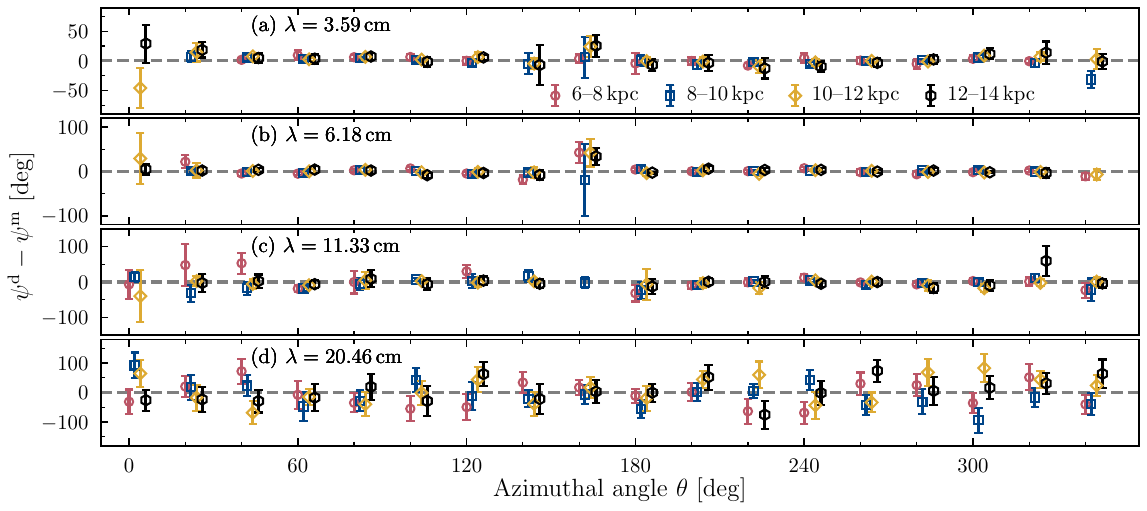}
\caption{The difference $\psi^\text{d} - \psi^\text{m}$ of the observed and fitted polarization angles at the wavelengths 3.59, 6.18, 11.33 and $20.46\cm$ (from top to bottom) across the four radial rings for the parameters given in \cref{tab:params_mcmc}: 6--8\kpc\ (red circles), 8--10\kpc\ (blue squares), 10--12\kpc\ (yellow diamonds) and 12--14\kpc\ (black hexagons).}
\label{fig_res_all}
\end{figure*}

The quality of the best-fitting model is further illustrated in Fig.~\ref{fig_res_all}  which shows the residuals $\psi^\text{d} - \psi^\text{m}$ for the best-fitting model of \cref{tab:params_mcmc}. In most cases, the residuals do not exceed the $1\sigma$ uncertainty of the observed polarization angles and do not exhibit any systematic trends, indicating a good statistical fit.

\begin{figure*}
\centering
\includegraphics[width=0.95\linewidth]{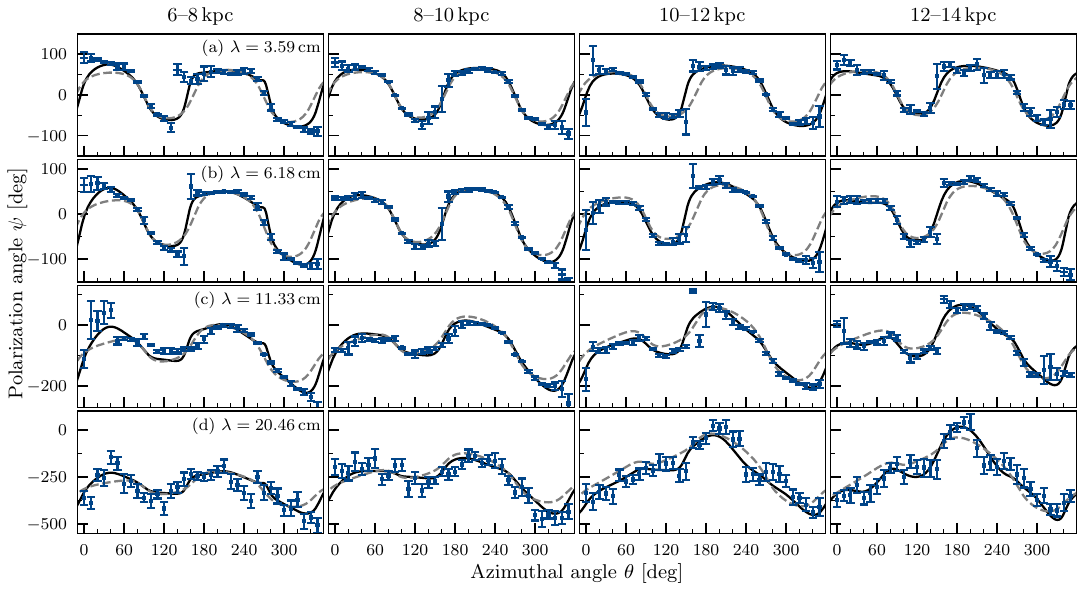}
\caption{As Fig.~\ref{fig:model} but for the observed polarization angles (symbols with $1\sigma$ error bars) averaged over narrower sectors of 10\deg\ in width. The results from the best-fitting model of \cref{tab:params_mcmc} are shown with black solid lines, whereas the dashed grey lines represent the contribution of the axisymmetric mode alone.}
\label{fig:model_10deg}
\end{figure*}

\subsection{Sector width}\label{SW}
Our results are obtained using the polarization angles averaged over sectors of 20\deg\ in width, large enough to reduce the contribution of the random magnetic field and to have a sufficient number  (at least five) of independent measurements in each sector (the relation of the beam size to the sector width is clear from Fig.~\ref{fig:M31_data1}).
To test the robustness of the best-fitting model of the magnetic field, we compare the fitted polarization angles with those averaged over narrower, 10\deg-wide sectors in Fig.~\ref{fig:model_10deg}. Although we do not use these data for the fitting, the model of \cref{tab:params_mcmc} is remarkably close to them. In particular, this data representation confirms the rapid changes of $\psi$ near the major axis ($\theta\approx0\deg,180\deg$) at the shorter wavelengths which are reproduced by the higher-order magnetic modes. The data points excluded from the analysis (shown in red in Fig.~\ref{fig:model}) are deviant in Fig.~\ref{fig:model_10deg} as well.

\begin{table*} 
\centering 
\caption{The mean values and $1\sigma$ uncertainties of the model parameters obtained by fitting the polarization angles at only the two shorter wavelengths, $\lambda\lambda 3.59$ $6.18\cm$ with $\xi_3= \xi_6 = 1$.}
\begin{tabular}{ccccccc}  
\toprule 
&Parameter& Unit & \multicolumn{4}{c}{Radial range [kpc]}\\ 
&& & \multicolumn{1}{c}{6--8} & \multicolumn{1}{c}{8--10} & \multicolumn{1}{c}{10--12} & \multicolumn{1}{c}{12--14} \\ 
\midrule 
    &$\RM_\text{fg}$ &rad\,m$^{-2}$ &$-121\pm11$  &$-99\pm7$  &$-92\pm9$  &$-96\pm13$\\ 
\midrule 
\multirow{2}{*}{$m=0$}  &$R_0$&rad\,m$^{-2}$ &$152\pm29$  &$176\pm25$  &$215\pm38$  &$204\pm37$\\ 
&$p_0$  &deg &$-16\pm4$  &$-14\pm3$  &$-14\pm3$  &$-22\pm4$\\ 
\midrule 
\multirow{3}{*}{$m=1$}  &$R_1$  &rad\,m$^{-2}$ &$49\pm29$  &$69\pm33$  &$137\pm31$  &$70\pm48$\\ 
&$p_1$  &deg &$-6\pm40$  &$2\pm22$  &$10\pm11$  &$12\pm33$\\ 
&$\beta_1$  &deg &$28\pm53$  &$13\pm24$  &$27\pm9$  &$-2\pm 42$\\ 
\midrule 
\multirow{3}{*}{$m=2$}&$R_2$&rad\,m$^{-2}$ &$54\pm13$  &$52\pm14$  &$38\pm22$  &$110\pm60$\\ 
&$p_2$  &deg &$49\pm21$  &$43\pm21$  &$-61\pm33$  &$-40\pm13$\\ 
&$\beta_2$  &deg &$-44\pm7$  &$-46\pm7$  &$41\pm16$  &$20\pm12$\\ 
\midrule 
\multirow{3}{*}{$m=3$}  &$R_3$  &rad\,m$^{-2}$ &$93\pm19$  &$60\pm19$  &$47\pm20$  &$79\pm47$\\ 
&$p_3$  &deg &$11\pm11$  &$10\pm16$  &$39\pm29$  &$14\pm27$\\ 
&$\beta_3$  &deg &$-25\pm3$  &$-31\pm7$  &$-23\pm8$  &$-41\pm8$\\
\midrule 
\multirow{3}{*}{$m=4$}  &$R_4$  &rad\,m$^{-2}$ &--  &--  &--  &$78\pm32$\\ 
&$p_4$  &deg &--  &-- &--  &$6\pm16$\\ 
&$\beta_4$  &deg &--  &--  &--  &$-46\pm 5$\\
\bottomrule 
\end{tabular} 
\label{tab:params_mcmc_3_6}
\end{table*} 

\subsection{The effects of depolarization}\label{Dep}

The mechanism of depolarization of the radio emission appears to be rather complicated and not quite certain but its correct modelling is essential for our purposes because depolarization makes M31 not transparent to the polarized emission at $\lambda\lambda11.33$ and $20.46\cm$, also making the magneto-ionic layer asymmetric along the line of sight which affects the relation of RM to the gas parameters. It is difficult to quantify these effects reliably, if possible at all, so we summarily represent them by introducing the factors $\xi_{11}$ and $\xi_{20}$. In order to confirm that the resulting model does not introduce any artefacts, we preformed the analysis using the two shorter wavelengths where Faraday depolarization is negligible. The results are shown in \cref{tab:params_mcmc_3_6} in the format similar to that of \cref{tab:params_mcmc}. The mean values of each parameter are close to those in \cref{tab:params_mcmc} and agree with them within the uncertainties, including the trends with the galactocentric distance. This confirms that the method to allow for depolarization suggested here is viable and efficient. The values of $\xi_{11}$ and $\xi_{20}$ derived (\cref{tab:params_mcmc}) can be used to explore the depolarization mechanisms. The advantages of using a larger number of wavelengths in the analysis are evident as this reduces uncertainties in the model parameters.

\begin{figure*}
    \centering
    \includegraphics[width=0.8\textwidth]{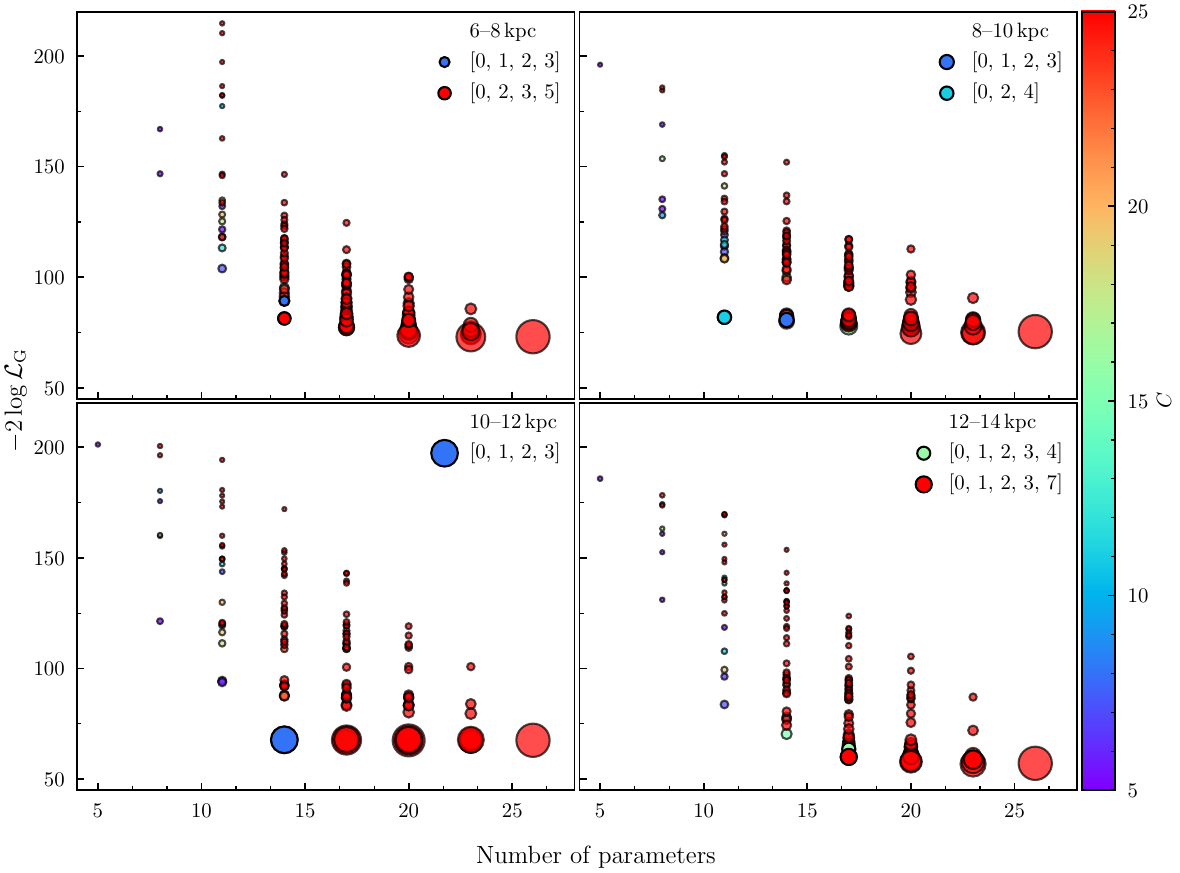} 
    \caption{The relationship between $-2\log\mathcal{L}_{\rm G}$ and the number of model parameters for any arbitrary combination of Fourier modes. For 20\deg-wide sectors, there are 128 mode combinations possible if we restrict the number of Fourier modes to $n_m=8$. The dominant $m=0$ mode is common to all the mode combinations. Each data point is colour-coded with the value of the Bayesian complexity parameter $\mathcal{C}$. The size of each data point (representing the likelihood value) is inversely proportional to difference of the likelihood from the reference likelihood value obtained by fitting all eight modes together. Thus, bigger symbols represent models with a larger likelihood.  The mode combination which provides the maximum likelihood (or the minimum of $-2\log\mathcal{L}_{\rm G}$) with a smaller number of model parameters (or the value of $\mathcal{C}$) is preferred by the observational data. In the legends, we show both the preferred mode combination from these comparisons in the lower entry and from our main analysis in the upper entry for the three rings where they differ; for $r=10\text{--}12\kpc$ they agree.}    
\label{apdx:complexity_plot}
\end{figure*}

\subsection{Various mode combinations}\label{appendix1}
In this section, we briefly discuss the variety of  magnetic mode combinations which also fit the observed polarization data. In \cref{tab:params_mcmc} and Fig.~\ref{fig:complexity_plot}, we only show a single Fourier mode combination for each radial range. To confirm that these are indeed the best models, we show in Fig.~\ref{apdx:complexity_plot} the Bayesian complexity $\mathcal{C}$ and the model likelihood for all 128 combinations of eight Fourier modes. For all the radial ranges, the Bayesian complexity increases as the number of model parameters and $-2\log\mathcal{L}_{\rm G}$ increases and then saturates after a certain combination. The best statistical combination for each number of azimuthal modes $N$ is given in the legend box of each ring.

\begin{itemize}
\item For the 6--8 kpc ring, the mode combinations $m=0,2,3,5$ and $0,1,2,3$ give equally good likelihood value with an equal number of free parameters in the model. Similarly, the mode combinations $0,1,2,3,4$ and $0,1,2,3,7$ give equally good likelihood values for the 10--12 kpc ring with an equal number of free parameters. The Bayesian complexity parameter gives no strong preference to either of the two mode combinations based on the existing data. We prefer the $m=0,1,2,3$ combination because it has a slightly smaller complexity parameter.

\item For the 8--10 kpc ring, the mode combination $0,2,4$ is preferred over  $0,1,2,3$ as it has fewer free parameters  and hence is less complex. The likelihood values are similar for both mode combinations. We prefer the mode combination $0,1,2,3$ because it is the same as the Fourier mode combination in the neighbouring rings.

\item For the 10--12 kpc ring, the mode combination $m=0,1,2,3$ is clearly preferred among the 128 possible mode combinations.
\end{itemize}

\begin{figure*}
    \centering
    \includegraphics[width=\textwidth]{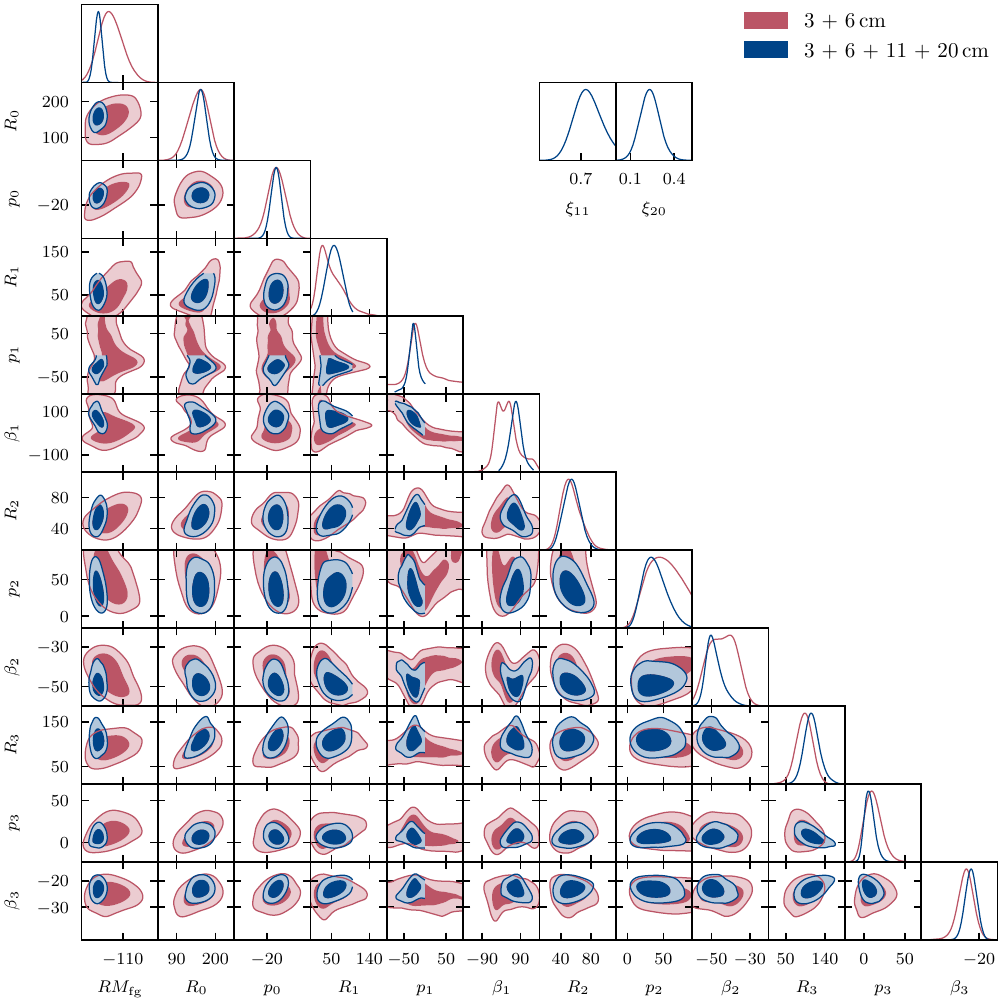}
    \caption{Same as Fig.~\ref{fig:mcmc_triangular_2}, but for the radial ring 6--8\,kpc.}
    \label{fig:mcmc_triangular_1}
\end{figure*}

\begin{figure*}
    \centering
    \includegraphics[width=\textwidth]{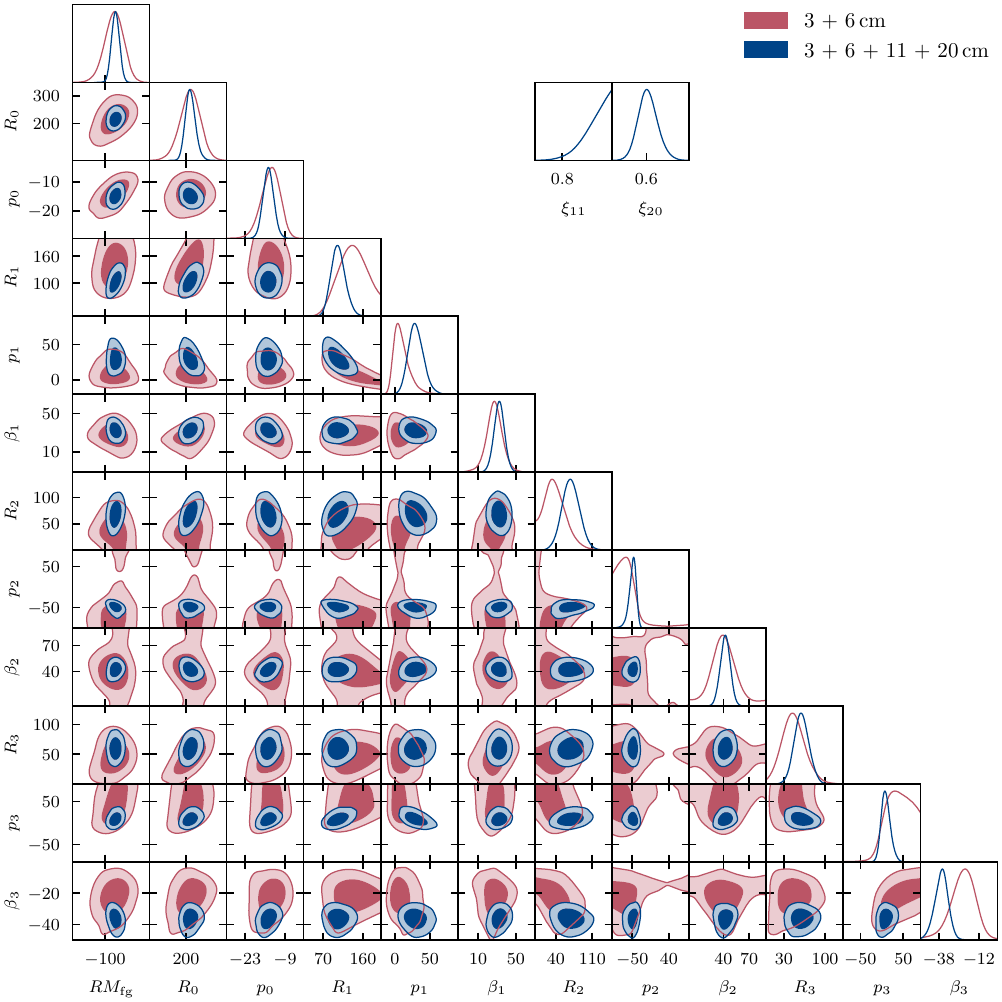}
    \caption{Same as Fig.~\ref{fig:mcmc_triangular_2}, but for the radial ring 10--12\,kpc.}
    \label{fig:mcmc_triangular_3}
\end{figure*}

\begin{figure*}
    \centering
    \includegraphics[width=\textwidth]{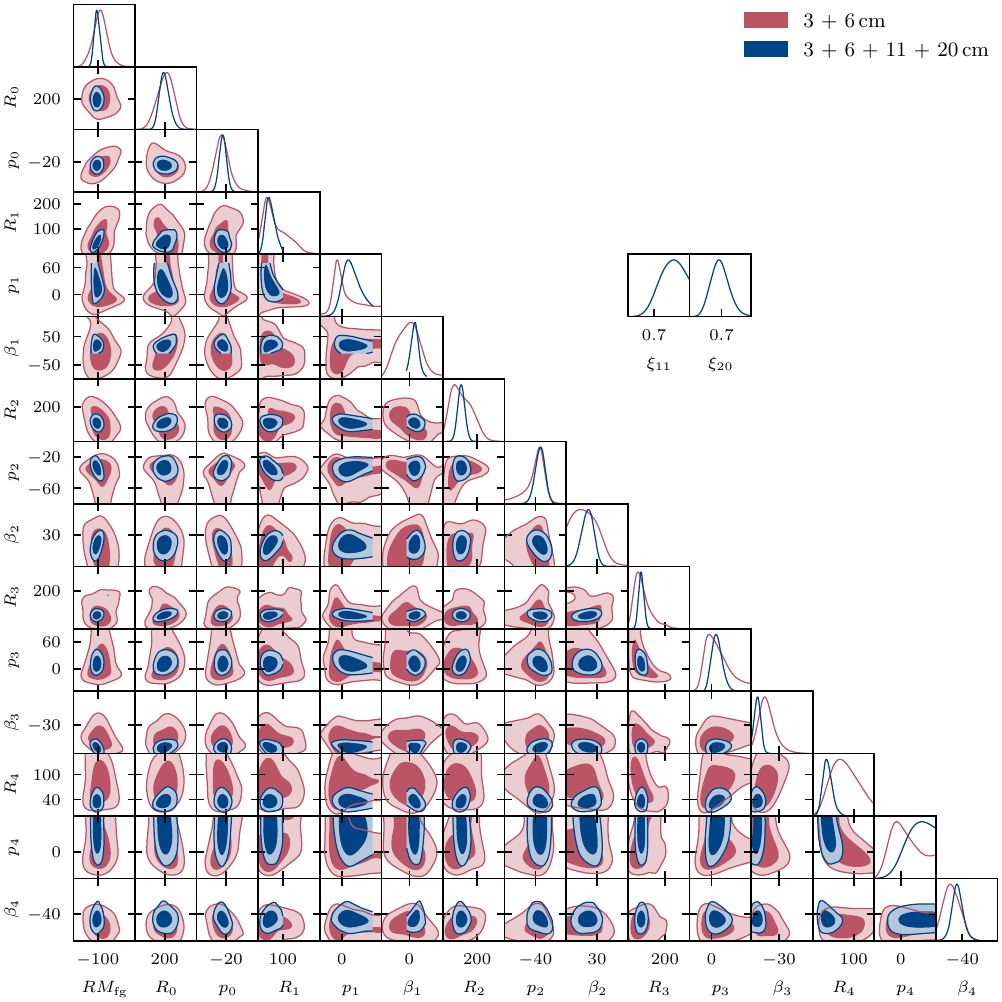}
    \caption{Same as Fig.~\ref{fig:mcmc_triangular_2}, but for the radial ring 12--14\,kpc.}
    \label{fig:mcmc_triangular_4}
\end{figure*}
\bsp	
\label{lastpage}
\end{document}